\documentclass[3p]{elsarticle}

\usepackage[english]{babel}

\usepackage{amsmath}
\usepackage{amssymb}
\usepackage{mathrsfs}
\usepackage{stmaryrd}
\usepackage[usenames]{color}
\usepackage[normalem]{ulem}
\usepackage{soul}

\usepackage{tensind}
\tensordelimiter{?}
\tensorformat{lrc}

\newcommand{\dd}{\mathrm{d}}

\newcommand{\ie}{\textit{i.e.}}
\newcommand{\eg}{\textit{e.g.}}

\title{A New Spherical Harmonics Scheme for Multi-Dimensional
  Radiation Transport I: Static Matter Configurations.}

\author[aei]{David Radice}
\ead{david.radice@aei.mpg.de}
\author[tapir]{Ernazar Abdikamalov}
\author[aei,lsu]{Luciano Rezzolla}
\author[tapir]{Christian D. Ott}

\address[aei]{Max Planck Institute f\"ur Gravitationsphysik, Albert
Einstein Institute, Potsdam, Germany}
\address[tapir]{TAPIR, California Institute of Technology, Pasadena,
CA}
\address[lsu]{Department of Physics and Astronomy, Louisiana State University,
  Baton Rouge, Louisiana, USA}

\begin{document}

\begin{abstract}
  Recent work by McClarren \& Hauck~\cite{McClarren10} suggests that the
  filtered spherical harmonics method represents an efficient, robust,
  and accurate method for radiation transport, at least in the
  two-dimensional (2D) case. We extend their work to the
  three-dimensional (3D) case and find that all of the advantages of the
  filtering approach identified in 2D are present also in the 3D case. We
  reformulate the filter operation in a way that is independent of the
  timestep and of the spatial discretization. We also explore different
  second- and fourth-order filters and find that the second-order ones
  yield significantly better results. Overall, our findings suggest that
  the filtered spherical harmonics approach represents a very promising
  method for 3D radiation transport calculations.
\end{abstract}
\maketitle

\section{Introduction}
\label{sec:intro}

Many phenomena in the universe involve the transport of radiation and
need to be modeled with radiation-transport techniques that are as
accurate as possible to maximize the match with observations. Examples
are nova and supernova explosions, gamma-ray bursts, star or planet
formation, luminous blue variable outbursts, stellar winds, etc. In these
examples, radiation plays a major role in exchanging energy and/or
momentum between different parts of the system. In most of these cases,
the radiation is composed of photons, but the radiation can also be
composed of neutrinos in studies of core-collapse supernova explosion
mechanisms {(see, \eg, ~\cite{Bethe:90,Janka07})} or in modeling the
torus orbiting the black hole produced in the merger of neutron stars
(see, \eg,~\cite{Ruffert98a}).

One of the main difficulties in solving the radiation-transport equation
arises from the multidimensionality of the problem.  Radiation is
described not only by the position of the radiation carriers, but also by
their direction of propagation and energy (or, equivalently, by their
momentum), making the problem $(6+1)$-dimensional in the most general
case (3 dimensions for the spatial coordinates, 2 for the angular
direction, 1 for the energy, and 1 for the time). Another source of
complication stems from the fact that many problems consist of regions of
strongly-varying optical depth. For example, many astrophysical systems
contain a central source of radiation, where the optical depth is high,
surrounded by more transparent outer regions. Due to high opacity near
the center, radiation migrates out of that region mostly via diffusion,
while in the outer parts, it streams to infinity more freely without much
interaction with matter. Correspondingly, in the limit of infinite
optical depth, the transport equation acquires a parabolic character,
while it maintains a hyperbolic one in all other regimes {(see, \eg,
  ~\cite{Mihalas84})}. Radiation-transport approaches must therefore
handle accurately not only these two distinct regimes, but also the
transition between the two, which is generally the most difficult to
treat. These features make the solution of the transport equation both
complex and computationally expensive.

There are two commonly used ways of simplifying the solution of the
transport equation. One approach consists in reducing the number of
degrees of freedom by assuming spherical or axial symmetry. While this is
a good simplification for some problems, there are many other situations
in which the system does not posses any spatial symmetries and hence the
transport equations need to be solved in three spatial dimensions.
Another way of simplifying the problem is via approximating the form of
the transport equation (this is equivalent to reducing dimensionality in
the momentum space). One of the simplest examples is the diffusion
approximation, where one approximates the transport equation with the
diffusion equation~(\eg, \cite{Pomraning:73}). This makes the equation
far simpler and computationally less expensive to solve~(\eg,
\cite{Mihalas84}). However, there is a price to pay for this: Although
the diffusion approximation is accurate at high optical depth, it leads
to incorrect results in low optical depth, \ie~``transparent'', regions.
This can be improved by using flux limiters (\eg, ~\cite{Castor:04}), but
the diffusion approximation still cannot correctly capture the anisotropy
of radiation commonly encountered in those regions {(see, \eg,
~\cite{Ott08})}.  Moreover, the treatment of semi-transparent regimes is
somewhat artificial because the radiation fluxes in those regions are
usually calculated via interpolation between the two fluxes calculated
assuming free-streaming and diffusive transport. There are several more
accurate ways of approximating the transport~(\eg, two-moment schemes
with analytic closures, etc.; \eg,~\cite{Brunner01}), but the solution of
the full (6+1)-dimensional transport equation is often not only
desirable, but actually necessary for an accurate modeling.

One of the most commonly used methods for solving the transport equation
in the multidimensional (both in space and momentum) case is the
discrete-ordinate ($S_N$) method, which solves the transport equation
along several directions in each spatial
zone~\cite{Castor:04,Ott08,Sumiyoshi:12,Godoy:12}. Unfortunately, this
method has several drawbacks. Most importantly, it suffers from ``ray
effects'' in multidimensional calculations~\citep{Morel:03}. Due to the
discrete nature of the angular representation, in fact, radiation cannot
reach regions between these discrete directions, leading to large spatial
oscillations in the transport variables. In addition, time-implicit $S_N$
methods require very complex solutions and parallelization
procedures~\citep{Adams:02}.

Monte-Carlo methods~(\eg, \cite{Fleck:71,Gentile:09,Abdikamalov12}) are
often regarded as the most accurate method for radiation transport, but
they are also not without drawbacks: Monte-Carlo solutions exhibit
statistical noise due to the finite sampling of the phase space. Since
this noise decreases only as $N^{-1/2}$, where $N$ is the number of Monte
Carlo particles, it can take many particles to produce a sufficiently
smooth solution, making large simulations computationally very expensive.

Another approach to radiation transport is the spherical harmonics, or
$P_N$, method. This method is based on an expansion of the radiation
intensity (or distribution function) in angles using spherical harmonics
basis functions. This results in a hyperbolic system of partial
differential equations for the expansion coefficients, which represent
angular moments in this basis. The spherical harmonics method has several
interesting characteristics. Due to hyperbolicity, the $P_N$ equations
approximate radiation as a series of waves with velocity bounded by the
speed of light~\cite{McClarren2008b}. This restriction is consistent with
the transport equation, in contrast to diffusion methods where radiation
propagates at infinite velocity. Moreover, the spherical harmonics
expansion exhibits formal spectral convergence to the true solution. Such
an expansion also preserves the rotational invariance of the solution,
unlike $S_N$ methods, where the absence of such invariance leads to the
ray effects mentioned above.\footnote{Here, rotational invariance means
that the operators of angular discretization and of rotation in space are
commutative. In other words, the result of any rotation and then of a
spherical harmonics angular discretization is the same as that of an
angular discretization and then of a rotation~\cite{Boyd00}. In the $S_N$
method, this is true only for those rotations which map the angular grid
onto itself.} Another advantage of the $P_N$ method is that it generally
uses less memory to model a given angular distribution at a given
accuracy compared to, \eg, the $S_N$ method. A $P_N$ truncation is
roughly equivalent to a $S_{N+1}$ solution, but, while the former has
$(N+1)^2$ degrees of freedom, the latter has $2(N+1)^2$, thus it requires
roughly twice as much memory. Given that memory requirements represent
one of the main difficulties in 3D radiation-transport calculations, a
factor two in memory saving represents a significant advantage.

However, spherical harmonics methods also have some negative aspects.
Most importantly, in transparent regions, the solution to the $P_N$
hyperbolic system exhibits non-physical oscillations. These oscillations
are related to the so-called Gibbs phenomenon that occurs when non-smooth
functions are approximated with smooth basis functions~\cite{Boyd00}. The
worst consequence of such oscillations is that they can cause the
radiation intensity to become negative, which may lead to negative matter
temperatures when the radiation transport is coupled to the matter energy
equation~\cite{McClarren2008b,Olson09}. There have been several attempts
to address this
problem~\cite{Olson00,Olson09,Brunner01,McClarren10,Olson12,Hauck2010}.
One of the most efficient, robust, and accurate approaches is the one by
McClarren \& Hauck~\cite{McClarren10}, in which filters are proposed to
remove oscillations of the radiation intensity. By filtering out the
oscillations, McClarren \& Hauck~\cite{McClarren10} were able to avoid
negative solutions while maintaining high angular accuracy. Their
filtering approach also has the advantage of being easily extendable to
high-order $P_N$ expansions, producing formal convergence to the
transport solution and preserving the equilibrium diffusion limit.

Although the results of~\cite{McClarren10} strongly suggest the idea of
applying a filter to spherical harmonics expansions is an efficient,
robust, and accurate way of doing $P_N$ radiation transport, there remain
several open questions. For example,~\cite{McClarren10} considered only
one type of filter, the so-called spherical-spline
expansion~\cite{Boyd00}. There are other types of filters that have some
interesting theoretical and numerical properties~\cite{Boyd00}, which
might be a more optimal choice for application to $P_N$ transport.
Moreover, McClarren \& Hauck~\cite{McClarren10} presented results only
for the 2D case, leaving open the question of how well their filtering
approach performs in 3D. Also, this filtering scheme, as realized
by~\cite{McClarren10}, and as we will discuss in detail later in this
paper, does not have a clear continuum limit as the spatial resolution
and/or timestep approaches zero, which implies that the solution cannot
be studied for spatial convergence.

In this paper, we reconsider the $P_N$ scheme and extend the work by
McClarren \& Hauck~\cite{McClarren10} in a number of ways. Firstly, we
reformulate the filtering procedure in such a way that it acquires a
clear continuum limit. Secondly, we investigate a wider range of filter
types and strength parameters. Finally, we perform calculations both in
2D and 3D. For this we have we have developed the new radiation transport
code {\tt Charon}, whose ultimate goal is that of performing 3D, general
relativistic multi-energy, multi-angle and velocity-dependent
radiation-transport calculations. In this paper, we present the first
step towards this goal.

{\tt Charon} uses the semi-implicit scheme of McClarren et
al.~\cite{McClarren2008a} for time integration. In this scheme, the
streaming parts are treated explicitly with a second-order Runge-Kutta
method, while the matter-coupling terms are treated implicitly because
of the stiffness introduced by the coupling. Also, the implicit system
of~\cite{McClarren2008a} is local to each spatial element, and for
linearized matter-coupling terms this implicit integration scheme
becomes trivial~\cite{McClarren2008a}.

Although the timestep in the semi-implicit approach is limited by the
Courant-Friedrichs-Lewy (CFL) condition based on the speed of light, this
is not a serious drawback for the kind of applications that we have in
mind. In radiation-hydrodynamics simulations, the timestep size would
still be limited by the dependence of the matter properties (\eg,
opacities and emissivities) on the matter temperature (and electron
fraction, if we are dealing with neutrinos with conserved lepton
number). Moreover, the semi-implicit approach is relatively easy to
parallelize (\eg, via domain decomposition) and has lower communication
requirements compared to fully implicit schemes. This should translate to
significant advantages in massively-parallel large-scale
radiation-hydrodynamics simulations. Moreover, in many relativistic
systems, the sound speed of the fluid is comparable to the speed of
light, and thus the radiation and hydrodynamics timescales are
comparable, reducing the extra cost in an explicit treatment of the
radiation streaming. The spatial discretization in {\tt Charon} is based
on the asymptotic-preserving linear discontinuous Galerkin (DG) scheme
\cite{Lowrie2002, McClarren2008a}.

This is the first in a series of three publications, with the present
one dealing with static matter configurations and focusing exclusively
on the transport methodology. For this reason, we do not consider the
coupling with hydrodynamical equations. As a result, when an
absorption occurs, the particles are simply removed from the system
and do not increase the matter temperature. The second paper will
include velocity-dependence and coupling to hydrodynamics, whereas in
the subsequent publication we will present a fully
general-relativistic algorithm.

The paper is organized as follows. In
Section~\ref{sec:boltsmann_equation}, we introduce the concept of
radiation distribution function and the relativistic Boltzmann
equation. In Section~\ref{sec:scheme}, we describe the numerical
schemes used in our code for frequency
(Section~\ref{sec:scheme_freq}), angular (Section~\ref{sec:ang_dis}),
spatial (Section~\ref{sec:spatial_discretization}), and time
discretization (Section~\ref{sec:scheme_time}). In
Section~\ref{sec:tests}, we present numerical tests of these
schemes. Finally, we summarize our results and provide our conclusions
in Section~\ref{sec:conclusion}. We use a spacetime signature
$(-,+,+,+)$, with Greek indices running from $0$ to $3$ and the Latin
indices from $1$ to $3$. We also employ the standard convention for
the summation over repeated indices.

\section{The relativistic Boltzmann equation}
\label{sec:boltsmann_equation}
\subsection{The distribution function for radiation}

Radiation is usually described in terms of the specific radiation
intensity, $I$, defined such that
\begin{equation}
\label{eq:specific.intensity}
  \dd \mathscr{E} = I\, \cos\theta\, \dd A \, \dd \nu\, \dd \Omega\, \dd
  t\,,
\end{equation}
represents the energy of radiation in frequency range $\dd\nu$
centered around $\nu$, traveling in direction $\Omega$ confined to a
solid angle element $\dd\Omega$, which crosses, within time interval $\dd
t$, an area $\dd A$ oriented such that $\theta$ is the angle between
the normal to the surface $\dd A$ and direction
$\Omega$ (\eg,~\cite{Pomraning:73}). In the case of neutrino transport
and/or in the relativistic case, it is more convenient to work
directly with the distribution function, $F$, which gives the density
of radiation carriers on a given point in phase space. The reason for
this is that (1) the distribution function is a Lorentz-invariant
quantity~\cite{Mihalas84}, and (2), as we discuss in more detail
later, the distribution function allows us to compute the number
density and the energy density of the radiation in a more natural way.

In order to introduce the distribution function, we first discuss the
notion of single-particle phase space in special relativity. In this
picture, the particles are described in terms of their positions in
spacetime, $x^\mu$, and their momentum four-vector, $p^\mu$, as measured
in a fiducial inertial frame. Using the normalization condition for
timelike vectors, the four-momentum has only three independent
components, which can be expressed in terms of three spatial components,
$p^i$, or in terms of radiation frequency, $\nu$, and two angles
$(\theta,\phi)$ that describe the propagation
direction:\footnote{Assuming the radiation carriers to be massless, their
  energy and frequency are simply related as $\varepsilon = h\nu$.}
\begin{equation}\label{eq:momentum.4vector}
  p^\mu = \frac{h\nu}{c}\, (1,\, \cos \phi \sin \theta,\,
  \sin \phi \sin\theta,\, \cos \theta)\,.
\end{equation}
Since we wish to define the distribution function in terms of $p^i$,
or, equivalently, in terms of $\nu, \phi$ and $\theta$, we need to
construct a Lorentz-invariant volume element, $\dd \Pi$, over the
manifold of the allowed momentum four-vectors. This is accomplished
with the choice~\cite{Cercignani2002}
\begin{equation}\label{eq:volume.element}
  \dd \Pi = \frac{\dd p^1 \dd p^2 \dd p^3}{-p_0}
          = \frac{h^2\nu}{c^2}\, \dd \nu \,\dd \Omega\,.
\end{equation}
The distribution function is then defined in such a way that the
quantity
\begin{equation}\label{eq:distribution.function}
  \dd N = F\, p^\mu t_\mu \, \dd^3 x\, \dd \Pi
    = \frac{h^3 \nu^2}{c^2}\, F\, \dd^3 x\, \dd \nu\, \dd \Omega\,,
\end{equation}
is the total number density of radiation carriers in a spatial volume
element $\dd^3x$ and phase-space volume element $\dd\Pi$ with
trajectories traversing a $t=\mathrm{const}$ hypersurface with normal
$\vec{t} = \partial_t$. Here, $t_\mu$ is the $\mu$ covariant component of
the vector $\vec{t} = \partial_t$.\footnote{Note that $\dd^3 x$ is not a
Lorentz-invariant quantity, while $p^\mu t_\mu \dd^3 x$ is one.}

Since $\dd^3x = \dd A\, \cos\theta\, \dd t$ and the energy per particle
is given by $h \nu$, we have $\dd \mathscr{E} = h \nu\, \dd N$. Using
this and comparing equation (\ref{eq:specific.intensity}) with
(\ref{eq:distribution.function}), we obtain:
\begin{equation}\label{eq:intensity.vs.dfunc}
  I = \frac{h^4 \nu^3}{c^2}\, F\,.
\end{equation}
Note that since $\dd N$ is a scalar, $F$ is also a scalar
quantity.\footnote{Note that this relation slightly differs from the one
  frequently encountered in the neutrino-transport literature (\eg,
  \cite{Burrows2000}):
\begin{equation}
I = \frac{\varepsilon^3}{h^3c^2}\, gF = \frac{\nu^3}{c^2}\, gF\,,
\nonumber
\end{equation} where $g$
  is the statistical weight of the particles ($g=1$ for massless
  neutrinos, $g=2$ for photons) and $\varepsilon$ is the particle
  energy. This is due to three reasons: First, our specific intensity
  given by Eq.~(\ref{eq:specific.intensity}) is defined in terms of
  energy per frequency interval, in contrast to energy per energy
  interval, as used in the neutrino-transport literature. Moreover, our
  distribution function already contains the factor $g$, as can {be} seen
  from its relation to the total number density of radiation carriers
  given by Eq.~(\ref{eq:distribution.function}).  Finally we use the
  Lorentz-invariant volume element given by Eq.~(\ref{eq:volume.element})
  instead of $\dd^3 p$.}

\subsubsection{The relativistic Boltzmann equation}

The special-relativistic Boltzmann equation can be written
as~\cite{Mihalas84}
\begin{equation}\label{eq:relativistic.boltzmann}
  p^\mu \frac{\partial F}{\partial x^{\mu}} = \mathbb{C}[F]\,,
\end{equation}
where $\mathbb{C}$ is {the collisional term describing} the
interaction of radiation with matter, while the left-hand-side of the
equation describes the propagation of radiation. In order to compute
$\mathbb{C}$, we express it in terms of the absorption, emission and
scattering coefficients.  To do that, we start by considering the
evolution equation for the intensity of radiation~\cite{Pomraning:73},
\begin{equation}\label{eq:classical.radiative.transfer.1}
  \frac{1}{c} \frac{\partial I}{\partial t} + n^i \frac{\partial
  I}{\partial x^i} = \eta - \kappa I + \frac{\kappa_s}{4\pi}
  \int \frac{\nu}{\nu'} K(\nu', \vec{n}\,' \rightarrow \nu, \vec{n})\,
  I(\nu', \vec{n}\,')\, \dd \Omega'\, \dd \nu'\,,
\end{equation}
where $\eta$ represents the radiative emissivity of the matter, $\kappa$
is the total extinction coefficient and combines the absorption and
scattering coefficients\footnote{Hereafter, the absorption or scattering
extinction coefficients are defined as absorption or scattering
opacitites or inverse mean-free paths.} $\kappa_a$ and $\kappa_s$, \ie,
$\kappa = \kappa_a + \kappa_s$, and $K$ is the scattering kernel,
expressing the probability of scattering from a given angle and frequency
over to another angle and frequency~\cite{Pomraning:73}.

Using equations (\ref{eq:momentum.4vector}),
(\ref{eq:intensity.vs.dfunc}), and
(\ref{eq:classical.radiative.transfer.1}), it is easy to obtain an
equation for $F$ in terms of the above extinction coefficients:
\begin{equation}\label{eq:boltzmann.sources.mixed}
  p^\mu \frac{\partial F}{\partial x^\mu} = \frac{c^2}{h^3}
  \frac{\eta}{\nu^2} - h \nu \kappa F + \frac{h \nu \kappa_s}{4\pi}\int
  \bigg(\frac{\nu'}{\nu}\bigg)^2 K(\vec{p}\,'\rightarrow\vec{p})\,
  F(\vec{p}')\, \dd\nu'\, \dd \Omega'\,.
\end{equation}
Notice that, since $F$ is a scalar, so is $\mathbb{C}[F]$, thus we
find the classical result that $\eta/\nu^2$, and $\nu\kappa$ are
invariant quantities~\cite{Mihalas84}.

Our scheme is in principle able to handle any type of scattering kernel,
but for simplicity, here we will only consider the case of elastic
scattering, \eg, scattering in which the radiation energy does not
change. In this case, the scattering kernel can be expresses
as~\citep{Burrows2000},
\begin{equation}\label{eq:scattering.kernel}
  K(\nu',\vec{n}\,'\rightarrow\nu,\vec{n}) = [ 1 +
  \sigma_a\, \vec{n} \cdot \vec{n}\,']\, \delta(\nu - \nu')\,,
\end{equation}
where the scattering anisotropy is modeled using only one coefficient
$\sigma_a$.

\section{Description of the scheme}
\label{sec:scheme}

In general, the distribution function, $F$, is a function of 7
variables: the time and spatial coordinates, $x^\mu$, the frequency
$\nu$ and the angles of propagation $\varphi$ and $\theta$. These
variables are usually defined either in the Eulerian (inertial) frame
or in the comoving frame (\ie, a set of frames, each of which has a
velocity that instantaneously equals that of the matter element,
\eg,~\cite{Mihalas84,Hubeny07}). In the case of static matter, as the
one considered in this paper, these two frames are identical. In the
scheme implemented by the \texttt{Charon} code, the distribution
function is expanded in the spatial coordinates using the linear DG
basis and in the angular variables using spherical harmonics. The
frequency is treated using the multi-group approach. This yields a
large system of ordinary differential equations that is then evolved
in time using a semi-implicit time integrator. The details of the
discretization are discussed in this Section.

\subsection{Frequency discretization}
\label{sec:scheme_freq}

We consider the case in which the distribution function has compact
support in a frequency space given by the interval $[0,\nu_{\max}]$.
Although this is not strictly valid in the general case, radiation
usually has negligible contribution above some cut-off
frequency. Therefore, in many practical applications, one can choose
$\nu_{\max}$ to be sufficiently large so that there is little radiation
beyond this frequency. For simplicity of illustration, we introduce a
uniform grid in this frequency space as $\nu_n = n\, \Delta \nu$, $n = 0,
1, \ldots, N_\nu+1$, where $\Delta \nu = \nu_{\max} / (N_\nu+1)$ (an
extension to a non-equidistant grid is conceptually trivial). The
associated intervals $[\nu_n, \nu_{n+1}]$ are commonly called frequency
or energy groups.  Using these groups, we can construct an orthonormal
basis $\{\chi_n\}_{n=0}^{N_\nu}$ as
\begin{align}\label{eq:energy.basis}
  \chi_n(\nu) &=
  \begin{cases}
    {1}/{\sqrt{V_n}}, & \textrm{if } \nu \in [\nu_n, \nu_{n+1}]\,, \\ 0,
    & \textrm{otherwise}\,,
  \end{cases}\,,&
  V_n &= \int_{\nu_n}^{\nu_{n+1}} h^3 \nu^2\, \dd \nu
      = \frac{h^3}{3} (\nu_{n+1}^3 - \nu_n^3)\,.
\end{align}
We then expand a function $f \in L^1(0,\nu_{\max})$ on this basis as
(for clarity we report the summation symbols in the expressions below)
\begin{align}
  f_{N_\nu}(\nu) &= \sum_{n=0}^{N_\nu} f^n \chi_n(\nu)\,,&
  f^n &= \frac{1}{\sqrt{V_n}} \int_{\nu_n}^{\nu_{n+1}} f(\nu)\, h^3\,
    \nu^2\, \dd \nu\,.
\end{align}
The truncated expansion, $f_{N_\nu}$, is then a first-order accurate (in
$L^1-$norm) approximation of $f$. We point out that, thanks to
our choice of basis (\ref{eq:energy.basis}), the final expansion of the
distribution function will involve integrals performed with respect to
the volume element in (\ref{eq:volume.element}). This allows us to ensure
exact conservation of the number of radiation particles in the
numerical treatment of the transport equation. Also, this choice of
the basis can be easily generalized to the case of curved spacetimes
(which will be the subject of our future work), where ensuring
conservation of radiation particles is particularly involved
\citep{Cardall2003}.

\subsection{Angular discretization}
\label{sec:ang_dis}

As orthonormal basis on the unit 2-sphere, $S_1$, we consider the real
spherical harmonics, $Y_{\ell m}$ (see \ref{appendix:sharmonics}), whose
orthonormality conditions are
\[
  \int_{S_1} Y_{\ell m}(\varphi,\theta)\, Y_{\ell' m'}(\varphi,\theta)
  \, \dd \Omega = \delta_{m m'} \delta_{\ell\ell'}\,.
\]
As a result, any function $f \in L^2(S_1)$ can be expanded in
spherical harmonics as
\begin{align}
  f_N(\varphi,\theta) &= \sum_{\ell = 0}^{N} \sum_{m = -\ell}^{\ell}
  f^{\ell m}\, Y_{\ell m}(\varphi, \theta)\,, & f^{\ell m} &= \int_{S_1}
  f(\varphi, \theta)\, Y^{\ell m}(\varphi,\theta)\, \dd \Omega\,,
\end{align}
where we have used the notation $Y^{\ell m}$ to denote the complex
conjugate of $Y_{\ell m}$ (which is equal to $Y_{\ell m}$ since we are
working with real spherical harmonics). If $f$ is a smooth function,
$f_N$ will converge to $f$ with spectral accuracy in the
$L^2-$norm~\cite{Boyd00}.

\subsection{The multi-group $P_N$ scheme}
We consider the following ansatz for the expansion of the distribution
function:
\begin{equation}\label{eq:solution.ansatz}
  F(x^\alpha, \nu, \varphi, \theta) =
  \sum_{n=0}^{N_\nu} \sum_{\ell = 0}^{N} \sum_{m = -\ell}^\ell
  F^{n\ell m}(x^\alpha)\, \chi_{n}(\nu)\, Y_{\ell m}(\varphi,\theta)\,,
\end{equation}
where
\[
  F_{n \ell m}(x^\alpha) = \int_{0}^{\infty} h^3 \nu^2\, \dd \nu
  \int_{\mathcal{S}_1} \dd \Omega\, F(x^\alpha, \nu, \varphi, \theta)\,
  Y_{\ell m}(\varphi, \theta) \, \chi_{n}(\nu)\,.
\]
To simplify the notation, we introduce the multi-index $A =
\{n,\ell,m\}$, and the basis functions,
\[
  \Psi_A(\nu, \varphi, \theta) \equiv \chi_{n}(\nu)\, Y_{\ell
  m}(\varphi,\theta)\,,
\]
so that Eq.~(\ref{eq:solution.ansatz}) becomes
\begin{equation}\label{eq:solution.ansatz.short}
  F(x^\alpha, \epsilon, \varphi, \theta) = \sum_A F^A(x^\alpha) \Psi_A(\epsilon,
  \varphi, \theta) = F^A \Psi_A\,.
\end{equation}
Note that the space spanned by our basis $\{\Psi_A\}$ is a vector space.
We adopt the usual convention of denoting vector components with an upper
index and co-vector components with a lower index. Linear operators
acting on vectors and co-vectors will have upper and lower indices
associated with their decomposition on an appropriate tensor product
combination of the canonical basis $\{\Psi_A\}$ and its dual
$\{\Psi^A\}$, defined by the requirement that $\int \Psi^A \Psi_B p^0 \dd
\Pi = ?[c]\delta^A_B?$. Note that, thanks to the orthornormality of real
spherical harmonics $\Psi^A = \Psi_A$.

Inserting Eq.~(\ref{eq:solution.ansatz.short}) into
Eq.~(\ref{eq:relativistic.boltzmann}), we obtain:
\begin{equation}\label{eq:scheme.derivation.step1}
  p^0 \frac{\partial F^B}{\partial t} \Psi_B + p^k \frac{\partial
  F^B}{\partial x^k} \Psi_B = \mathbb{C}[F]\,.
\end{equation}
Multiplying Eq.~(\ref{eq:scheme.derivation.step1}) by $\Psi^A$ and
integrating with respect to $\dd \Pi$, we then obtain
\begin{equation}\label{eq:charon.scheme}
  \frac{\partial F^A}{\partial t} + ?[c]{\mathcal{P}}^k^A_B?
  \frac{\partial F^B}{\partial x^k} = \mathbb{S}^A[F]\,,
\end{equation}
where we have again used Eq.~(\ref{eq:momentum.4vector}) and exploited the
orthonormality of the basis. We have also defined
\begin{equation}\label{eq:charon.stiff}
  ?[c]{\mathcal{P}}^k^A_B? \equiv \int p^k\, \Psi^A\, \Psi_B\, \dd \Pi\,,
\end{equation}
and
\begin{equation}\label{eq:charon.source}
  \mathbb{S}^A[F] \equiv \int \mathbb{C}[F]\, \Psi^A\, \dd \Pi\,.
\end{equation}
The coefficients (\ref{eq:charon.stiff}) can be computed exactly using a
quadrature formula of high-enough order (see \linebreak
\ref{appendix:sharmonics}). Since they are independent of position and
time, they can be pre-computed and stored for later usage.

The spectral decomposition of $?[c]{\mathcal{P}}^k^A_B?$, which
determines the behavior of Eq.~(\ref{eq:charon.scheme}), is well
known, see \eg~\cite{Brunner2005}. In particular, it has been shown
that the eigenvalues are strictly bounded by the speed of light
$c$. While this implies that there is no superluminal propagation of
radiation, it leads to slower-than-light motion of radiation waves for
finite $N$ (the radiation velocity converges to the correct value with
increasing $N$)~\cite{Brunner01}. This is particularly evident in
low-order $P_N$ free-streaming solutions. For instance, the maximum
propagation speed for $P_1$ is $c/\sqrt{3}$. Filtering can also affect
the propagation velocity of radiation \cite{McClarren10}. In the
multidimensional case, $?[c]{\mathcal{P}}^k^A_B?$ has also zero-speed
modes that have to be treated carefully in Godunov-based schemes to
avoid numerical instabilities \cite{Brunner2005}.

In the general case, the source term~(\ref{eq:charon.source}) has to
be computed at run-time, but for the particular case of a source term
in the form (\ref{eq:scattering.kernel}), and assuming that the
opacity coefficients are constant in each of the energy groups (as
commonly done in multi-group schemes \cite{Pomraning:73}), the source
terms can be pre-computed up to constant factors. Under these
assumptions, the source term becomes
\begin{equation}
  \mathbb{S}^A[F] = e^A - \kappa_n F^A - \kappa_{s,n} [
  \perp^A_{\phantom{A}B} - \sigma_{a,n} ?[c]\Delta^A_B?] F^B =
  e^A + ?[c]S^A_B? F^B\,,
\end{equation}
where\footnote{Notice that the term in square brackets is the number
  of emitted particles.}
\[
  e^A \equiv \int \bigg[\frac{c^2 \eta}{h^4 \nu^3}\bigg]\,
    \Psi^A(\nu, \varphi, \theta)\,  h^3 \nu^2\, \dd \nu\, \dd \Omega\,,
\]
$\perp^A_{\phantom{A}B}$ is the projector perpendicular to $Y_{00}$,
\[
  \perp^A_{\phantom{A}B} \equiv  ?[c]\delta^A_B? -
    ?[c]\delta^A_{n00}? ?[c]\delta^{n00}_B?\,,
\]
and $?[c]\Delta^A_B?$ is the anisotropy matrix:
\[
  ?[c]\Delta^A_B? \equiv \frac{1}{4\pi} \int h^3 \nu^2\, \dd \nu\, \dd
  \Omega\, \Psi^A(\nu, \varphi, \theta) \int \Psi_B(\nu, \varphi',
  \theta')\, \vec{n}\,'\cdot\vec{n}\, \dd \Omega'\,,
\]
which can also be pre-computed. Note that we have denoted the opacity
coefficients in the energy group $[\nu_n, \nu_{n+1}]$ with the
subscript $n$.

\subsection{Spatial discretization}
\label{sec:spatial_discretization}

\begin{figure}
  \begin{center}
    \includegraphics[scale=0.7]{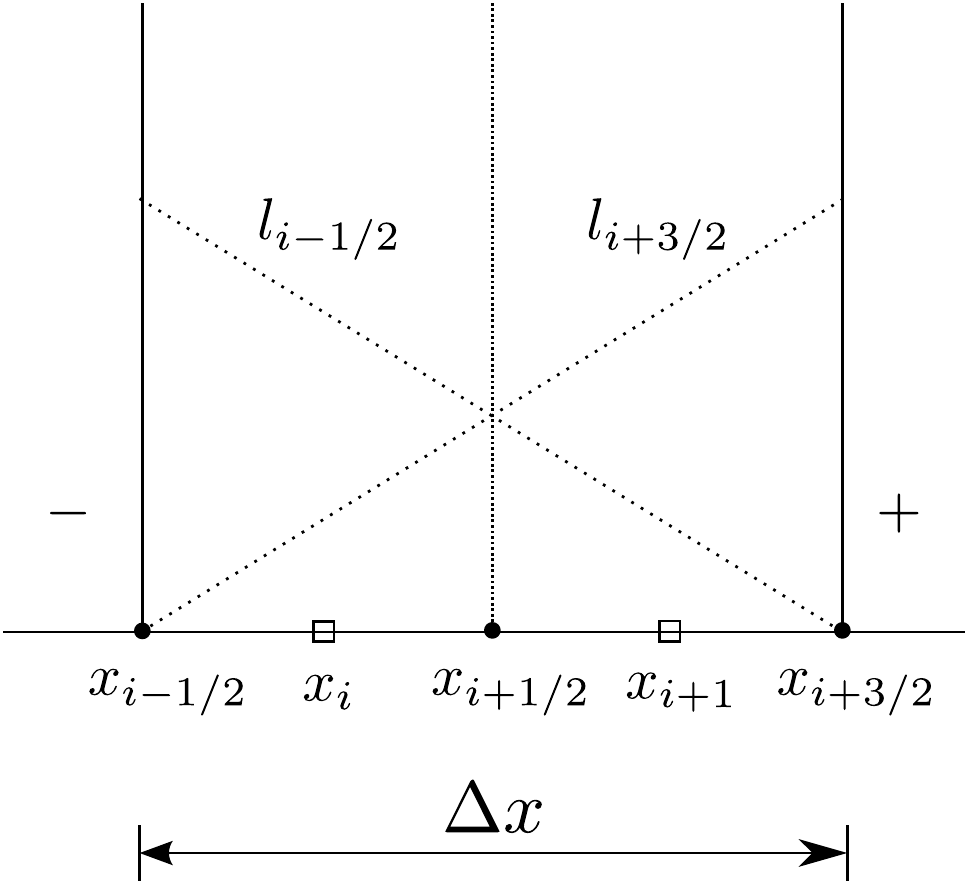}
  \end{center}
  \caption{\label{fig:ldg.grid}The grid structure for the spatial
    linear discontinuous Galerkin discretization. The white squares
    represent the cell centers, while the black dots show the cell
    interfaces. The dotted lines show the values of the Lagrangian
    basis $l_{i-1/2}$, $l_{i+3/2}$.  Finally the $+$ and $-$ show the
    interfaces where the inter-element fluxes, $\mathcal{F}^{\pm}$,
    see (\ref{eq:dg_fluxes}), are computed.}
\end{figure}

We discretize the system of Eqs.~(\ref{eq:charon.scheme}) in space using
the asymptotic-preserving (AP) linear DG scheme, see
\eg~\cite{Lowrie2002, McClarren2008a}. The scheme we present here is
restricted to the case of orthogonal grids, but we point out that
discontinuous Galerkin methods can be (and have been) extended to
general, unstructured grids \cite{cockburn_2001_rkd}. We recall that a
scheme is AP if it reproduces a discretization of the diffusion limit of
the continuum transport equation in the limit of small mean-free-path.
This is an important property since it guarantees that the diffusion of
radiation has a correct rate even if the mean-free-path is small compared
to the spatial grid size. If a scheme is not asymptotically preserving,
then the diffusion rate becomes unphysical when the mean-free-path is
unresolved.

For simplicity of notation, we consider a simplified 1D version of
Eq.~(\ref{eq:charon.scheme}):
\[
  \frac{\partial F^A}{\partial t} +
  ?[c]{\mathcal{P}}^{1}^A_B? \frac{\partial F^B}{\partial x} = 0\,,
\]
where the multidimensional case will be discussed at the end of this
Section. Furthermore, we employ a uniform numerical grid $x_i = i
\delta x$, while extension to a non-uniform grid is conceptually
trivial. We then choose the points of the grid that are used to
construct elements of width $\Delta x = 2 \delta x$, as shown in
Fig.~\ref{fig:ldg.grid}. This rather uncommon grid structure has been
adopted to ease the integration of \texttt{Charon} with existing
general-relativistic hydrodynamic codes that use traditional
finite-volume schemes.

In the classical linear DG scheme, the degrees of freedom are usually
identified with the one-sided limits of the solution at the points
$x_{i-1/2}$ and $x_{i+3/2}$ from the interior of the element, while in
our case we evolve the cell-centered values at $x_i$ and
$x_{i+1}$. The semi-discrete equations for the evolution of these
cell-centered values can be easily obtained, since any function $u$ in
the finite-element space can be written as
\[
  u(x) = u_{i-1/2} l_{i-1/2}(x) + u_{i+3/2} l_{i+3/2}(x)\,,
\]
where
\begin{align}
  l_{i-1/2}(x) &= 1 - \frac{x - x_{i-1/2}}{x_{i+3/2} - x_{i-1/2}}\,,&
  l_{i+3/2}(x) &= \frac{x - x_{i-1/2}}{x_{i+3/2} - x_{i-1/2}}\,,
\end{align}
so that
\begin{align}
  u_{i-1/2} &= \frac{3}{2} u_{i} - \frac{1}{2} u_{i+1}\,, &
  u_{i+3/2} &= -\frac{1}{2} u_i + \frac{3}{2} u_{i+1} \,,
\end{align}
and
\begin{align}
  u_{i} &= \frac{3}{4} u_{i-1/2} + \frac{1}{4} u_{i+3/2}\,, &
  u_{i+1} &= \frac{1}{4} u_{i-1/2} + \frac{3}{4} u_{i+3/2}\,.
\end{align}
The resulting numerical scheme is
\[
  \Delta x \frac{\dd ?[c]F^A_i?}{\dd t} = ?[c]{\mathbb{F}}^A_i?\,,
\]
where the flux terms in the element $[x_{i-1/2}, x_{i+3/2}]$ are given by
\begin{align}
\label{eq:dg_fluxes}
  ?[c]{\mathbb{F}}^A_i? &\equiv \frac{3}{2} \mathcal{F}^- -
  \overline{\mathcal{F}} - \frac{1}{2} \mathcal{F}^+ \,,&
  ?[c]{\mathbb{F}}^A_{i+1}? &\equiv \frac{1}{2}\mathcal{F}^- +
    \overline{\mathcal{F}} - \frac{3}{2} \mathcal{F}^+\,.
\end{align}
Here,
\[
  \overline{\mathcal{F}} \equiv \frac{1}{2} \Big[
   \big(?[c]{\mathcal{P}}^{1}^A_B?\big)_i
      ?[c]F^B_i? +
   \big(?[c]{\mathcal{P}}^{1}^A_B?\big)_{i+1} ?[c]F^B_{i+1}?\Big]\,,
\]
is the average flux, while
\[
  \mathcal{F}^- \equiv \frac{1}{2}
  \bigg[
    ?[c]{\mathcal{P}}^{1}^A_B? \big(?[c]F^B_L? + ?[c]F^B_R?\big) -
    ?[c]{\mathcal{R}}^{1}^A_C?  \mathrm{max}(v, |?[c]\Lambda^{1}^C_D?|)
    ?[c]{\mathcal{L}}^{1}^D_B?  \big(?[c]F^B_R? - ?[c]F^B_L?\big)
  \bigg]\,,
\]
is the flux computed from the exact solution of the Riemann problem at
$x_{i-1/2}$ with the left ``$L$'' and right ``$R$'' states
$?[c]F^A_L?$ and $?[c]F^A_R?$. The term $\mathcal{F}^+$ is the flux
across $x_{i+3/2}$ and is calculated analogously to
$\mathcal{F}^-$. In the previous equations, we have decomposed
$?[c]{\mathcal{P}}^{1}^A_B?$ as
\[
  ?[c]{\mathcal{P}}^{1}^A_B? = ?[c]{\mathcal{R}}^{1}^A_C?
    ?[c]\Lambda^{1}^C_D? ?[c]{\mathcal{L}}^{1}^D_B?\,,
\]
where $?[c]{\mathcal{R}}^{1}^A_C?$ are the matrices of the right
eigenvectors, while $?[c]\Lambda^{1}^C_D?$ and
$?[c]{\mathcal{L}}^{1}^D_B?$ are the eigenvalues and the left
eigenvectors of $?[c]{\mathcal{P}}^{1}^A_B?$, respectively. We have
also introduced $v > 0$ which is taken to be the first positive
abscissa of the adopted Legendre quadrature and it is used to
introduce extra numerical dissipation on the zero speed modes
similarly to what is done in \cite{Brunner2005}. Notice again that the
spectral decomposition of $?[c]{\mathcal{P}}^1^A_B?$, discussed in the
previous Section, can also be pre-computed at the beginning of the
calculations.

Having described the numerical scheme for the 1D problem, we can
construct the multidimensional numerical scheme for
Eq.~(\ref{eq:charon.scheme}) by repeating the same construction in
every direction:
\begin{equation}\label{eq:ldg.full}
  \frac{\dd ?[c]F^A_{i,j,k}?}{\dd t} =
  e^A + ?[c]S^A_B? F^B +
  \frac{1}{\Delta x} ?[c]{\mathbb{F}}^A_{i,j,k}? +
  \frac{1}{\Delta y} ?[c]{\mathbb{G}}^A_{i,j,k}? +
  \frac{1}{\Delta z} ?[c]{\mathbb{H}}^A_{i,j,k}? \,,
\end{equation}
where the fluxes in the $y$ and $z$ direction, $\mathbb{G}$ and
$\mathbb{H}$, are computed analogously to the ones in the $x$ direction.

To avoid creation of false extrema in the numerical solution we use the
slope limiting technique~\cite{cockburn_2001_rkd}. Among the different
limiters that we have implemented are (1) the so-called ``step limiter'',
which simply reduces the scheme to a first order discontinuous Galerkin
scheme, (2) the ``minmod'' and (3) the asymptotic-preserving ``minmod2''
limiters \cite{McClarren2008}. The reason for using these particular
limiters is that they have been well tested in the context of the
transport equation (\eg, \cite{McClarren2008}). Furthermore, the minmod2
limiter has been studied in detail in the context of linear DG methods,
where it has been shown that it does not affect smooth solutions away
from local extrema~\cite{Cockburn1991}, thus yielding a scheme with a
very small dissipation.

\subsection{Time discretization}
\label{sec:scheme_time}

For the time integration, we use the predictor-corrector method
proposed by McClarren~et~al.~\cite{McClarren2008a}. In this approach,
the streaming terms that model the transport of radiation are treated
explicitly, while the source terms responsible for interaction with
matter are treated implicitly. The use of this particular time
integrator is motivated by the fact that this yields a relatively
inexpensive, stable and asymptotic-preserving scheme. As discussed in
Section~\ref{sec:intro}, the fact that the streaming terms are treated
explicitly makes this scheme particularly easy to parallelize, while
the associated CFL constraint is not particularly demanding for
applications involving fluid moving at relativistic velocities and
general-relativistic gravity.

In order to simplify the notation, we rewrite Eq.~(\ref{eq:ldg.full})
as
\begin{equation}\label{eq:ldg.short}
  \frac{\dd F^A}{\dd t} = e^A + ?[c]S^A_B? F^B + \mathcal{A}^A[F]\,,
\end{equation}
where $\mathcal{A}^A[F]$ is a shorthand for the treatment of the spatial
flux terms. For the time integration of Eq.~(\ref{eq:ldg.short}), we use
the following two-step semi-implicit asymptotic-preserving scheme. Given
the solution $?[c]F^A_k?$ at time $k \Delta t$, we first perform a
predictor step
\[
  \frac{?[c]F^A_{k+1/2}? -
  ?[c]F^A_k?}{\Delta t/2} = \mathcal{A}^A[F_k] + ?[c]e^A_k? + ?[c]S^A_B?
  ?[c]F^B_{k+1/2}?\,\,,
\]
to obtain the solution at time $(k+1/2)\Delta t$ and then a corrector
step:
\[
  \frac{?[c]F^A_{k+1}? -
  ?[c]F^A_k?}{\Delta t} = \mathcal{A}^A[F_{k+1/2}] + ?[c]e^A_{k+1/2}? +
  ?[c]S^A_B?  ?[c]F^B_{k+1}?\,,
\]
to obtain the solution at time $(k+1)\Delta t$. At both stages, the
absorption, emission and scattering terms are treated implicitly, while
the streaming terms are treated explicitly. The explicit part of this
scheme is second-order accurate in time, while its implicit part is
first-order accurate~\cite{McClarren2008a}.

\subsection{Filtering}
\label{sec:scheme_filter}

{Filtering is a common procedure to reduce the effects of the Gibbs
  phenomenon in spectral methods for numerical solution of partial
  differential equations \cite{CHQZ88,Radice2011}}. Filtered spherical
harmonics expansions have been successfully used in, \eg, meteorology
(see \eg~\cite{Boyd00} and references therein) and the effects of
filtering on the truncation error of a spectral expansion are now
reasonably well understood \cite{Vandeven1991}. However, the use of
filters to mitigate (and, in most situations, remove) the occurrence
of negative solutions in $P_N$ schemes has only been proposed recently
by McClarren and Hauck~\cite{McClarren10}.

In their work, the authors propose to filter the spherical harmonic
expansion of the solution after each timestep using a spherical-spline
filter. Applying this suggestion to the spherical harmonic expansion
of $F$ we obtain (for clarity we report here the summation symbols)
\begin{equation}
  \big[\mathscr{F}(F)\big](\varphi, \theta) = \sum_{\ell=0}^N \sum_{m =
  -\ell}^\ell \left[\frac{1}{1 + \alpha \ell^2 (1 + \ell^2)}\right]
  F^{\ell m} Y_{\ell m}(\varphi, \theta)\,,
\end{equation}
where
\begin{equation}
  \alpha \equiv \frac{c\Delta t}{\Delta x} \frac{1}{N^2}
    \frac{1}{(\sigma_t L + N)^2}\,,
\end{equation}
and $L$ is a characteristic length scale used to make $\alpha$
dimensionless, while $\sigma_t$ is chosen to be of the same order of
magnitude as $\kappa$.

The filtered spherical harmonics, $F\!P_N$, scheme has several
interesting properties. Filtering has been found to be very effective
and robust in removing numerical oscillations in $P_N$ solutions,
while preserving the rotational invariance of the scheme. Furthermore,
for this particular choice of $\alpha$, filtering turns off
automatically in the limit $N\to\infty$, thus it does not spoil the
convergence of the spherical harmonics expansion.

However, one important drawback, also remarked by McClarren and
Hauck~\cite{McClarren2008}, is that the filtered $P_N$ scheme does not
have a clear continuum limit as $\Delta x, \Delta t \to 0$. This is
unfortunate because it implies that the filtered $P_N$ scheme,
$F\!P_N$, cannot be interpreted as a system of partial differential
equations. This in turn means that the quality of the $F\!P_N$
solution will depend on the spatial grid resolution in a way that is
hard to predict. The ultimate and most important implication is that
an $F\!P_N$ solution cannot be studied for spatial convergence.

To solve this problem, we propose a modification/generalization of
the $F\!P_N$ scheme as follows. We introduce a strength parameter, $s
\geq 0$, to be specified later, and construct the filtered expansion
as
\begin{equation}
  \label{eq:filter_expansion}
  \big[\mathscr{F}(F)\big](\varphi, \theta) = \sum_{\ell=0}^N \sum_{m =
  -\ell}^\ell \bigg[\sigma\Big(\frac{\ell}{N+1}\Big)\bigg]^s F^{\ell m}
  Y_{\ell m}(\varphi, \theta)\,,
\end{equation}
where $\sigma(\eta)$ is a \emph{filter function} of order $p$, that
is, a function $\sigma\in C^p\big(\mathbb{R}^+;[0,1]\big)$ such
that\footnote{\label{foot:vandeven} Here we ignore the technical
  requirement for Vandeven's theorem that $\sigma^{(k)}(1) =
  0\ \textrm{for } k = 0,1,2,\ldots p-1$, which is not satisfied by
  our filters (nor by the one proposed by \cite{McClarren10}). This is
  a condition that does not influence the formal accuracy of the
  filtered expansion with respect to the unfiltered truncated
  expansion, but it is mainly needed to prove the convergence of the
  filtered expansion in the case in which the unfiltered expansion is
  not converging point-wise (for instance due to the presence of
  discontinuities)~\cite{Gottlieb1997,Tanner2006}.}
\begin{align}
  \sigma(0) &= 1\,, &
  \sigma^{(k)}(0) = 0\,, \ \textrm{for } k = 1,2, \ldots p-1\,.
\end{align}
Notice that, since the filter strength depends only on $\ell$, this does
not destruct the rotational invariance of the scheme.\footnote{This is a
consequence of the classical addition theorem for spherical harmonics
(\eg, \cite{Boyd00}).} Furthermore, as the order of the spherical
harmonics, $N$, increases, the effect of filtering automatically
decreases, so that the convergence of the scheme for $N\to\infty$ is
retained. More specifically, for a filter of order $p$, we expect a
convergence order of $\sim p - 1$, as suggested by Vandeven's theorem for
Fourier expansion~\cite{Vandeven1991}, see also
\cite{hesthaven_2008_fls}.

In our analysis we have considered two second-order and two
fourth-order filters. The first one is the classical second-order
Lanczos filter\footnote{Note that the Lanczos filter is usually
  defined as $\sigma(\eta) = \sin\pi\eta/\pi\eta$ to have a
  first-order zero at $\eta=1$, as discussed in footnote
  \ref{foot:vandeven}. Our modified Lanczos filter yields a more
  uniform damping of high-order modes and works very well in our
  experiments.}:
\begin{equation}
  \sigma_{\textrm{Lanczos}}(\eta) = \frac{\sin\eta}{\eta}\,,
\end{equation}
while the second and third choices are given by the
ErfcLog filter~\cite{Boyd1996},
\begin{equation}
  \sigma_{\textrm{ErfcLog}}(\eta) = \frac{1}{2} \mathrm{Erfc} \Bigg\{ 2
  p^{1/2} \bigg(|\eta|-\frac{1}{2}\bigg)
  \sqrt{\frac{-\log[1-4(\eta-1/2)^2]}{4(\eta-1/2)^2}} \Bigg\}\,,
\end{equation}
of order $p = 2,4$. Finally, the fourth filter is the fourth-order
spherical-spline filter
\begin{equation}
  \label{eq:spline_filter}
  \sigma_{\textrm{SSpline}}(\eta) = \frac{1}{1 + \eta^4}\,.
\end{equation}
We point out that, with our definition, the spherical-spline
filter~(\ref{eq:spline_filter}) is very similar to the one used in
\cite{McClarren10}, but is \emph{not} exactly equivalent. The reason for
using a slightly different filter form is that the filter
of~\cite{McClarren10} is not compatible with the form of
Eq.~(\ref{eq:filter_expansion}) as it cannot be written in terms of a
function $\sigma(\cdot)$ of $\ell/(N+1)$.

In addition, we have considered only even-order filters since the
truncation error associated with these filters can be interpreted as a
numerical viscosity of order higher than two \cite{meister_2009_sfd},
while for odd-order filters the leading truncation error is of the
dispersion type~\cite{meister_2009_sfd}. Moreover, we also do not
consider filters of orders higher than $4$. This is because, as we
will see later, the fourth-order filters are already too weak to
completely remove oscillations, suggesting that even higher order will
be even less efficient.

In our scheme, we filter the solution after each sub-step of the time
integrator. This yields the following scheme
\begin{align}
  \label{eq:filter.a}
  \frac{?[c]F^A_{\ast}? -
  ?[c]F^A_k?}{\Delta t/2} &= \mathcal{A}^A[F_k] + ?[c]e^A_k? + ?[c]S^A_B?
  ?[c]F^B_{k+1/2}?, \\
  \label{eq:filter.1}
  ?[c]F^A_{k+1/2}? &= ?[c]{\mathscr{F}}^A_B? ?[c]F^B_{\ast}?, \\
  \label{eq:filter.b}
   \frac{?[c]F^A_{\ast\ast}? -
  ?[c]F^A_k?}{\Delta t} &= \mathcal{A}^A[F_{k+1/2}] + ?[c]e^A_{k+1/2}? +
  ?[c]S^A_B?  ?[c]F^B_{k+1}?, \\
  \label{eq:filter.2}
   ?[c]F^A_{k+1}? &= ?[c]{\mathscr{F}}^A_B? ?[c]F^B_{\ast\ast}?,
\end{align}
where $?[c]{\mathscr{F}}^A_B?$ is a diagonal matrix representing the
filtering operation.

We should remark that both our scheme and the one
by~\cite{McClarren10}, cannot be interpreted as a continuum problem,
in the sense that the equations
(\ref{eq:filter.a})$-$(\ref{eq:filter.2}) do not, in general,
represent a discretization of any system of partial differential
equations. The main reason is that $?[c]{\mathscr{F}}^A_B?$ is not
idempotent, \ie, $?[c]{\mathscr{F}}^A_C?  ?[c]{\mathscr{F}}^C_B?  \neq
?[c]\delta^A_B?$, so that the filtering operations in
Eqs.~(\ref{eq:filter.1}) and (\ref{eq:filter.2}) do not have a
well-defined behavior in the limit $\Delta t \to 0$.  In the case in
which $?[c]{\mathscr{F}}^A_B?$ is idempotent, the scheme has indeed a
continuum limit, but it can be easily demonstrated that the $F\!P_N$
method is just the $P_M$ method for some $M \leq N$.\footnote{The
  reason is that the only {idempotent} filter is the cut-off filter,
  that is, the filter that simply sets to zero all the modes with
  $\ell > M$ for some $M$, while leaving unaffected all the modes with
  $\ell \leq M$.}

This problem can be solved by making the strength parameter $s$ (and
$?[c]{\mathscr{F}}^A_B?$ with it) depend on the timestep. In order to
see that, we consider the behavior of our scheme for a given mode $u =
F^{\ell m}$ with $\ell \neq 0$. Let $q = \sigma\big(\ell / (N+1)\big)$,
where $\sigma$ is a filter function. Then the effect of filtering on
$u$ in each of the two filtering steps in, \eg,
Eq.~(\ref{eq:filter.2}) is simply:
\[
  u_{k+1} = q^s u_{\ast\ast}\,.
\]
This can be rewritten as
\[
  \frac{u_{k+1} - u_{\ast\ast}}{\Delta t/2} = \frac{1}{\Delta t/2}
    [q^s - 1] u_{\ast\ast}\,.
\]
If we let $s = \beta \Delta t$, then, in the limit of $\Delta t \to
0$, we obtain
\begin{equation}\label{eq:filter.ode}
  \frac{\dd u}{\dd t} = \beta\, \log q\, u\,.
\end{equation}
In other words, we can interpret filtering as a first-order, operator
split, discretization of the system of equations
\begin{equation}\label{eq:filtered.pn}
  \frac{\partial F^A}{\partial t} + ?[c]{\mathcal{P}}^k^A_B?
  \frac{\partial F^B}{\partial x^k} = e^A + ?[c]S^A_B? F^B + \beta
  ?[c]L^A_B?  F^B\,,
\end{equation}
where $?[c]L^A_B?$ is a diagonal matrix with coefficients
$\log\sigma\big(\ell / (N+1)\big)$. This is the desired continuum
limit. The physical interpretation is that filtering is equivalent to
a forward-peaked scattering operator (notice that $\sigma(0) =
1$). Finally, we note that we can estimate the filter effective
opacity by looking at the dissipation rate for the highest-order
multiple of the expansion as
\[
  \sigma_{\rm eff} = - \beta \log \sigma\big(N/(N+1)\big)\,.
\]

\section{Tests}
\label{sec:tests}

In this Section, we present some tests of the numerical schemes described
above as implemented in our {\tt Charon} code. {\tt Charon} uses 3D
Cartesian coordinates in space and is currently parallelized employing
hybrid OpenMP/MPI parallelization using the domain decomposition
method. It uses the open-source {\tt Cactus Computational
  Toolkit}~\cite{Goodale2002,cactusweb}, which provides MPI
parallelization, input/output, and restart capabilities.

\subsection{1D diffusion of a step function}
\label{sec:1d_step_diff_test}

In this first test, we primarily focus on verifying the ability of our
scheme to handle diffusion of radiation when the opacity is high and the
mean-free-path is small compared to the grid spacing. In this limit, the
continuous hyperbolic transport equation displays parabolic character to
leading order~\cite{McClarren2008}. Despite this, there is no guarantee
that a numerical scheme for solving the hyperbolic system will be AP,
that is, will reproduce a valid discretization of the asymptotic limit of
the continuous equations (cf. {the} discussion in
Section~\ref{sec:spatial_discretization}).

Consider therefore the following initial conditions for the radiation
energy density $E \equiv \int I\, \dd \Omega\, \dd \nu,$ in a
non-moving infinite medium with a constant (isotropic and elastic)
scattering opacity:
\begin{equation}
\label{eq:1d_diff_id}
E(z, t = 0) = \mathrm{H}(z+1/2)\mathrm{H}(1/2-z)\,,
\end{equation}
where $\mathrm{H}(\cdot)$ is the Heaviside step function. If the
scattering opacity is high, the solution of the transport equation is
well approximated by the solution of the diffusion equation. The
corresponding diffusion equation has the following analytic solution
to problem~(\ref{eq:1d_diff_id})
\begin{equation}
E(z,t) = \frac{1}{2}\left[\mathrm{Erf}\left(
\frac{z+1/2}{2\sqrt{t/\tau}}\right) -
\mathrm{Erf}\left(\frac{z-1/2}{2\sqrt{t/\tau}}
\right)\right]\,,
\end{equation}
where $\mathrm{Erf(\cdot)}$ is the error function and $\tau \equiv
3\kappa_s/c$ is the diffusion timescale, where $\kappa_s$ is the total
scattering opacity, which we set $\kappa_s =
10^5$~\cite[\eg,][]{McClarren2008}.

We employ five different schemes for this test: the step scheme (\ie, a
DG scheme with step-limiter, or, equivalently a first-order FV scheme),
two linear DG methods employing minmod and minmod2 limiters and two
finite-volume (FV) methods also employing minmod and minmod2 limiters. In
our implementation, the finite-volume scheme is obtained from the linear
DG scheme by simply replacing the linear DG slope with the one obtained
from the reconstruction procedure.

In all of our runs, we use the $P_1$ scheme because in 1D there are no
negative solutions and thus filtering is not necessary, and because the
radiation is nearly isotropic in such a diffusive regime so that $P_1$
scheme should be sufficiently accurate. We perform calculations using
three different resolutions $\Delta z = 0.16, 0.08$ and $\Delta z =
0.04$, with the grid ranging from $-2$ to $2$, and imposing periodic
boundary conditions at the outer boundaries. We choose the CFL factor to
be $0.25$ and we recall that the maximum CFL factor that guarantees the
$L^2$-stability of our scheme is $1/3$ in 1D \cite{cockburn_2001_rkd}. In
all of our tests, the CFL factor is mainly chosen for convenience in
order to have a sufficient number of timesteps within a given time
interval. Moreover, in many radiation-transport calculations, in the
absence of hydrodynamical equations, the truncation error due to the time
discretization is expected to be small compared to other sources of error
(\eg, the angular and spatial discretization).

\begin{figure}
  \begin{center}
    \includegraphics{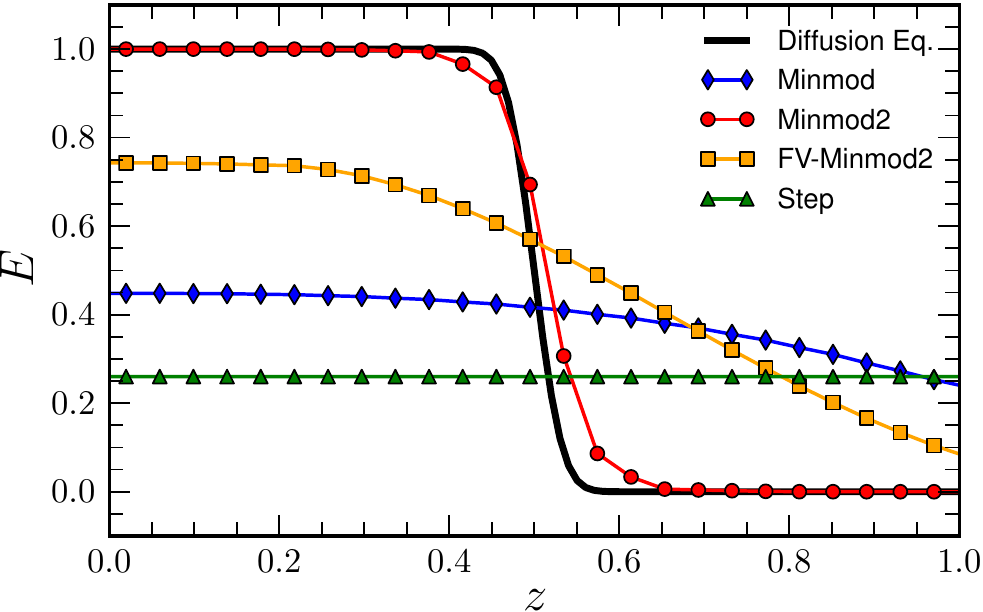}
  \end{center}
  \caption{Radiation energy density as a function of the $z$ coordinate
    at time $100/c$ for the 1D diffusion of the step
    function~(\ref{eq:1d_diff_id}). The thick black line represents the
    analytical solution of the corresponding diffusion equation, the line
    with red {circles} corresponds to the linear DG with minmod2 limiter
    solution, the line with orange {squares} represents the solution from
    the finite-volume scheme with minmod2 reconstruction, while the lines
    with blue diamonds and with green {triangles} show the results
    obtained with linear DG with minmod and step limiters, respectively.
    The symbols also mark the values of the numerical solution at each
    grid point (\ie, we show two points for each element).}
  \label{fig:step}
\end{figure}

Figure~\ref{fig:step} shows the radiation energy density as a function of
$z$ coordinate for the run with $\Delta z = 0.04$ at time $100/c$. The
thick black line corresponds to the analytic solution, while the other
lines show numerical results obtained with the above schemes. The line
with red circles corresponds to the linear DG with minmod2 limiter
solution. The line {with orange squares} represents the solution from the
finite-volume scheme with minmod2 reconstruction. Finally, the lines with
{blue diamonds} and green {triangles} show the results obtained with
linear DG with minmod and step limiters, respectively. Note that the
different symbols also mark the value of the numerical solution at each
grid point (\ie, we show two points for each element).

The linear DG method with minmod2 agrees well with the analytical result.
This is expected since this scheme has the correct asymptotic
limit~\cite{McClarren2008}. In contrast, all other schemes overestimate
the diffusion rate. In particular, the step scheme produces the worst
results. It reaches stationarity already at time $t /\tau \sim 10^{-4}$,
which is much smaller than the diffusion timescale for this problem. The
linear DG and finite-volume schemes with minmod yield identical results
and for this reason we show only the results from the linear DG scheme.
Both are only marginally better than the step algorithm.  These results
are in overall agreement with the ones reported by \cite{McClarren2008}
for a very similar test.

The FV scheme with the minmod2 reconstruction produces results that
are relatively accurate compared to the linear DG and finite-volume schemes
with the minmod limiter, even though the observed diffusion timescale
(at the current resolution) is still orders of magnitude larger than
the physical one. Finally, we point out that the results obtained
from the other lower-resolution runs (not shown here) are in overall
agreement with the ones presented here.

\subsection{1D diffusion of a sine wave}
\label{sec:1d_sine_diff_test}

Next, we explore convergence of the numerical solution to the asymptotic
one. To this scope we consider the 1D diffusion of a sine wave and thus
adopt as initial conditions the energy density given by
\begin{equation}
E(z,t = 0) = 3\sqrt{4 \pi} \left[\sin\left(\frac{\pi
z}{3}\right) + 1\right]\,,
\end{equation}
which has the following analytic solution in the diffusion limit
\begin{equation}
E(z,t) = 3\sqrt{4 \pi} \left[1 +
\exp\left(-\frac{\pi^2}{9}\, \frac{t}{\tau} \right)
  \sin\Big(\frac{\pi z}{3}\Big) \right]\,.
\end{equation}
For this test, our computational domain ranges from $-3$ to $3$ and we
use eight different resolutions ranging from $\Delta z= 0.3$ to $\Delta z
= 0.009375$. The CFL factor is chosen to be $0.3$

\begin{figure}
  \begin{center}
    \includegraphics{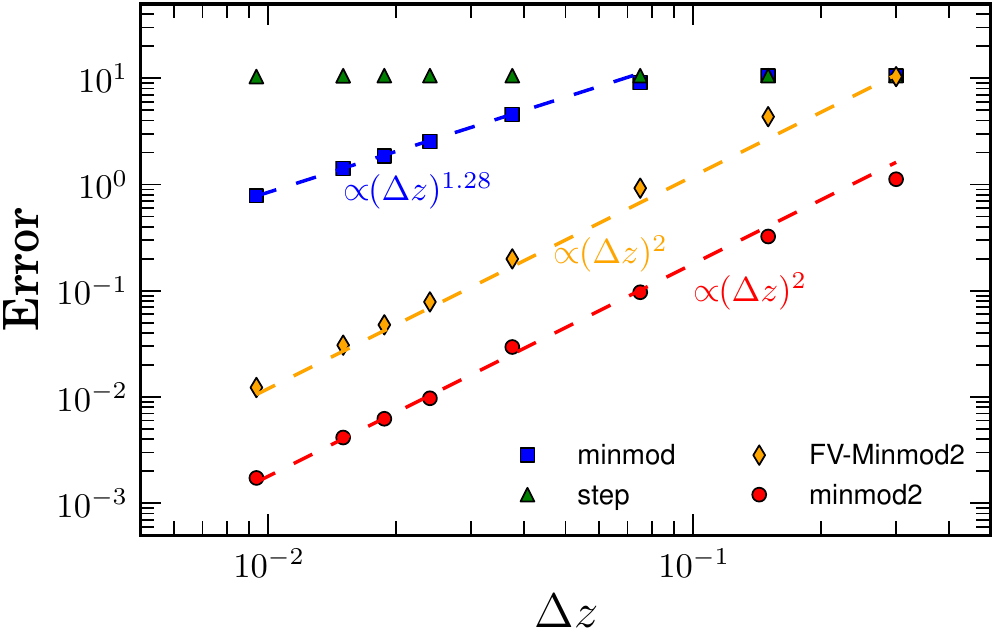}
  \end{center}
  \caption{$L^\infty-$norm of the deviation of the numerical result from
  the asymptotic solution as a function of numerical resolution for the
  diffusion problem of a sine function. The lines with red {circles} and
  blue {squares} correspond to the linear DG schemes with minmod2 and
  minmod limiters, respectively, while the line with orange {diamonds}
  represents the FV scheme with minmod2 reconstruction. Finally, the line
  with green triangles corresponds to the step scheme.}
  \label{fig:sine_diffusion}
\end{figure}

Figure~\ref{fig:sine_diffusion} shows the $L^\infty-$norm of the
deviation of the numerical result from the asymptotic solution as a
function of numerical resolution at time $t = 1000/c$. As expected, the
linear DG scheme with minmod2 (line with red {circles}) exhibits
approximately second-order convergence for the entire range of
resolutions shown in the plot, while the linear DG with minmod (line with
blue {squares}) starts converging only when $\Delta z \sim 10^{-1}$,
afterwards it converges with order $\simeq 1.28$. The step scheme (line
with green triangles) does not show any sign of convergence. These
results are again consistent with what was observed in
\cite{McClarren2008}. Finally, the finite-volume with minmod2
reconstruction (line with orange diamonds) exhibits second-order
convergence even though this scheme is not asymptotic preserving.

\begin{figure}
  \begin{minipage}{0.49\hsize}
    \includegraphics[width=\textwidth,height=0.3\textheight]{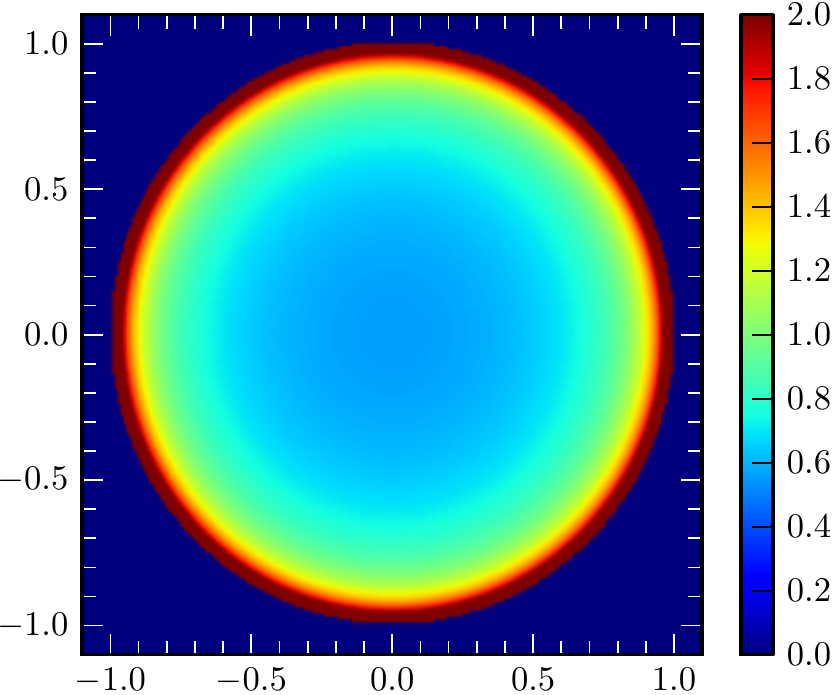}
    \center{\vskip -0.25cm (a) Analytic solution}
    \vskip 0.75cm
    \includegraphics[width=\textwidth,height=0.3\textheight]{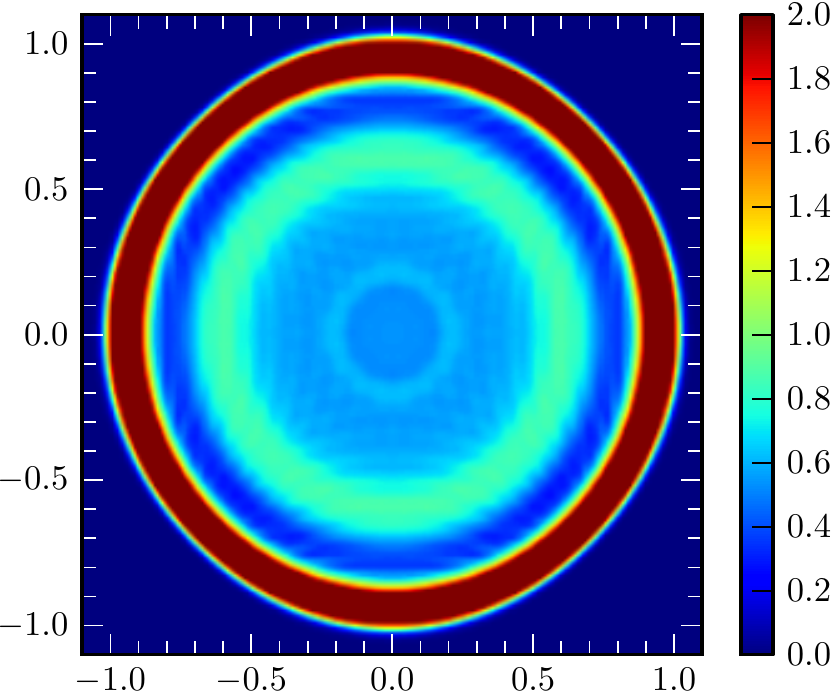}
    \center{\vskip -0.25cm (c) $F\!P_7$ with spherical-spline filter}
  \end{minipage}
  \hskip 0.5cm
  \begin{minipage}{0.49\hsize}
    \includegraphics[width=\textwidth,height=0.3\textheight]{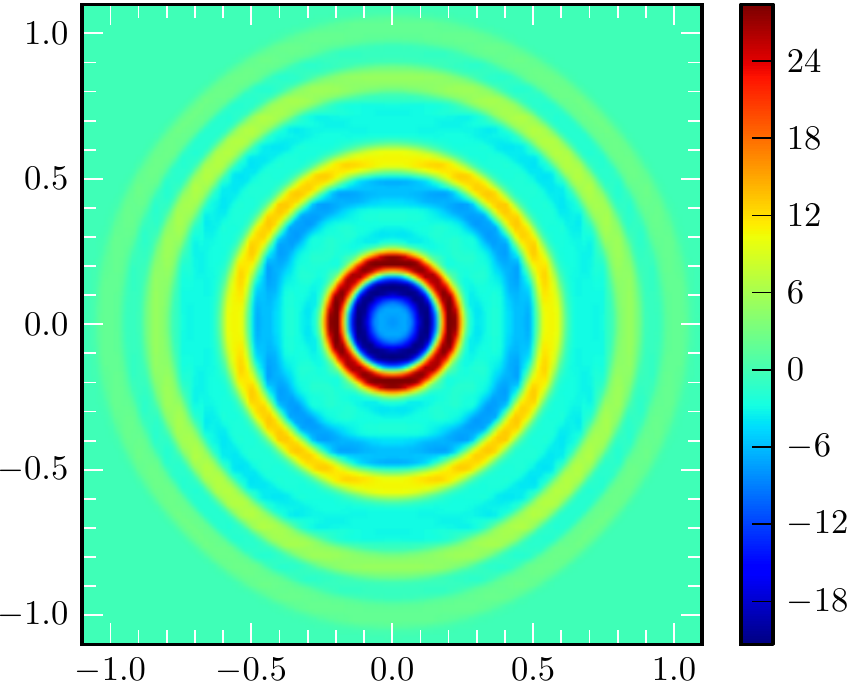}
    \center{\vskip -0.25cm (b) Original $P_7$}
    \vskip 0.75cm
    \includegraphics[width=\textwidth,height=0.3\textheight]{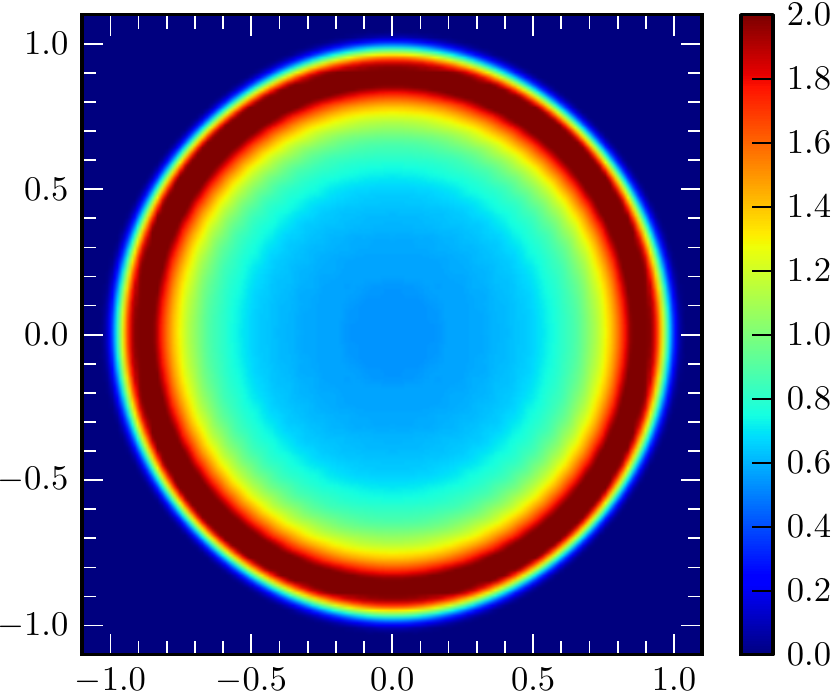}
    \center{\vskip -0.25cm (d) $F\!P_7$ with Lanczos filter}
  \end{minipage}
  \caption{Colormaps of the radiation energy density on the $x-y$
    plane at $t=1/c$ for the line problem with different methods. The
    upper left panel shows the analytic solution, while the upper
    right panel shows the pure $P_7$ solution (note the considerable
    difference in scale). The lower left panel shows the $F\!P_7$
    solution with spherical-spline filter with effective opacity
    $\sigma_{\rm eff} = 20$, while the lower right panel shows the
    $F\!P_7$ solution with the Lancsoz filter with the same effective
    opacity.}
  \label{fig:linesource2d}
\end{figure}

\subsection{The {2D} line-source problem}
\label{sec:line_problem}

As a first multidimensional problem used to benchmark different
implementations of the filtered spherical harmonics discretization
schemes we consider the so-called ``line-source'' problem, where we
have initial conditions given by\footnote{The initial conditions are
  3D but we exploit the cylindrical symmetry to solve the problem on
  the $(x,y)$ plane only.}
\begin{equation}
I(x,y,z,\Omega,t = 0) = \frac{E^0}{4\pi}\delta(x)\delta(y) \,,
\end{equation}
which represent an isotropic pulse of radiation with a total energy
$E^0$ concentrated along the $z$-axis in vacuum~\cite{Brunner02}. This
radiation field propagates in vacuum as it evolves in time according
to the following analytical solution
\begin{equation}
\label{eq:line_problem_solution}
E(x,y,t) =
\frac{E^0}{2\pi}\frac{\mathrm{H}(ct-r)}{ct\sqrt{c^2t^2-r^2}}\,,
\end{equation}
where $r \equiv \sqrt{x^2+y^2}$ is the distance from the $z$-axis and
$\mathrm{H}(\cdot)$ is the unit step function and $\delta(\cdot)$ is
the Dirac delta function. According to this solution, the radiation
field consists of a front that forms a cylindrical shell that travels
outwards at the speed of light, while in the interior, the radiation
energy density smoothly decreases along the radial direction.

We point out that, while this test is actually one-dimensional in
cylindrical coordinates, it becomes particularly challenging for
radiation-transport codes, except for Monte-Carlo codes, when solved on a
two-dimensional {Cartesian} grid (as we do). In these coordinates, the
radiation beam, which originates from a single spatial grid zone, has a
very forward-peaked distribution in angle. This is a huge challenge for
both spatial and angular discretization schemes.  Moreover, such a
configuration favors negative solutions in the $P_N$ scheme. Indeed, the
analytical $P_N$ solution to this problem was shown to have regions with
negative values of the energy~\cite{Brunner02, McClarren2008b,
McClarren10}, while the $P_1$ solution even exhibits a negative
singularity~\cite{McClarren08c}. For all the results presented here, we
use a grid with resolution $\Delta x = \Delta y = 0.02$ and a CFL factor
of $0.0625$. Furthermore, we choose $E_0 = \sqrt{4\pi}$.

Figure~\ref{fig:linesource2d} displays the colormap of the radiation
energy density in the $x-y$ plane at $t=1/c$. The upper left panel shows
the analytic solution, while the upper right panel shows the pure $P_7$
solution (note the considerable difference in scale). As expected, the
$P_7$ solution exhibits unphysical oscillations in the radial direction
that are absent in the analytical solution {to the full transport
problem}~(\ref{eq:line_problem_solution}). The lower left panel of
Fig.~\ref{fig:linesource2d} shows instead the $F\!P_7$ solution with
spherical-spline filter with effective opacity $\sigma_{\rm eff} = 20$
(the dependence of the solution on the order $N$ and on the value of the
filter strength $\sigma_{\rm eff}$ will be discusses below). In this
case, the radial oscillations are significantly reduced compared to the
unfiltered $P_7$ solution, similar to what was found
in~\cite{McClarren10}. Finally, the lower right panel shows the $F\!P_7$
solution with the Lancsoz filter with the same effective opacity. In this
case, the amount of oscillations is even smaller and we get a result that
is closer to the analytical one. The reason seems to be that the Lanczos
filter, being a second-order filter, is more effective in reducing the
appearance of oscillations. The solution obtained with the
spherical-spline filter is still characterized by the presence of a ring
structure that resembles the more oscillatory (unfiltered) $P_N$
solution. This structure does not disappear even for the values of the
filter strength as high as $\sigma_{\rm eff} = 10^4$, suggesting that
this is a result of a shortcoming of this particular filter. We point out
that we have repeated these runs with the second-order (fourth-order)
ErfcLog filter and we obtain a result similar to the one with the
second-order Lanczos (fourth-order ErfcLog) filter, suggesting that the
order of the filter plays the most important role, at least for this
test.

\begin{figure}
  \begin{center}
    \includegraphics{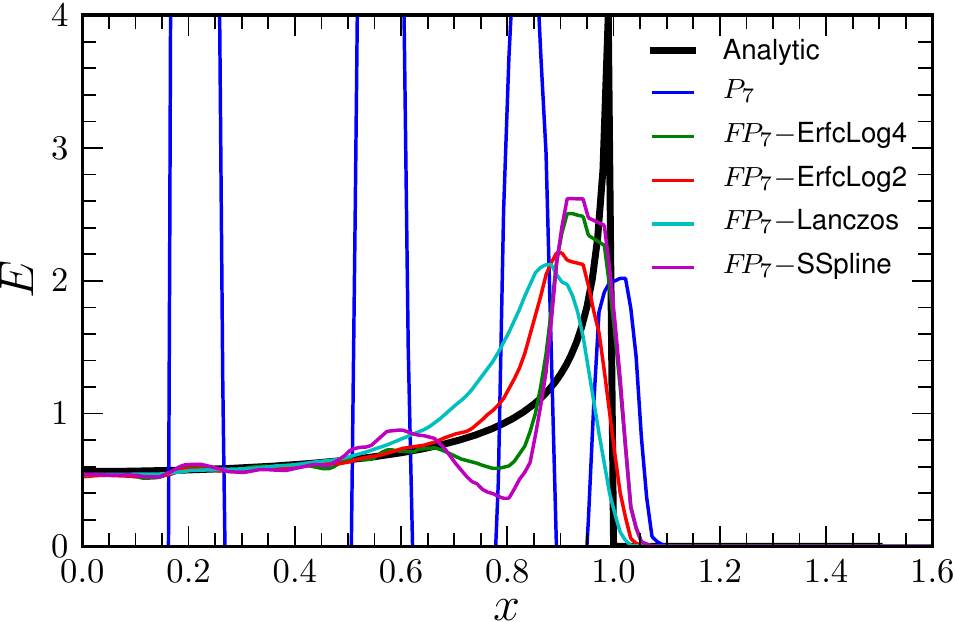}
  \end{center}
  \caption{The radiation energy density as a function of the $x$
  coordinate at $t=1/c$ for the line problem test. The {thick}
  black line corresponds to the analytic solution, while the rest of the
  lines represent the results from $P_7$ calculations without filter
  (blue line), with fourth-order ErfcLog filter (green line), with the
  second-order ErfcLog filter (red line), with the Lanczos filter (cyan
  line), and with the fourth-order spherical-spline filter (magenta
  line).}
  \label{fig:linesource1d_filters}
\end{figure}

A more quantitative measure of this test is shown in
Fig.~\ref{fig:linesource1d_filters}, where we plot a 1D cut of the
radiation energy density as a function of the $x$-coordinate at
$t=1/c$. The {thick} black line again corresponds to the analytic
solution, while the rest of the lines represent the results from $P_7$
calculations without filter (blue line), with the fourth-order ErfcLog
filter (green line), with the second-order ErfcLog filter (red line),
with the Lanczos filter (cyan line), and with the spherical-spline
filter (violet line). All of these runs with filters are performed
using a filter strength of $\sigma_{\rm eff}=20$.  We can easily
notice again the presence of large oscillations in the unfiltered
$P_7$ solution (which are larger than the scale of the plot).  The
fourth-order spherical-spline and ErfcLog filters are able to suppress
most of the oscillations and remove the negative values.
Nevertheless, both solutions are still affected by the oscillations
(although to a much smaller extent compared to the $P_7$
solution). The second-order Lacnzos and ErfcLog filters, on the other
hand, are able to remove most of the oscillations and give the best
numerical solutions, overall.

\begin{figure}
  \begin{center}
    \includegraphics[width=7.75cm,height=5.5cm]{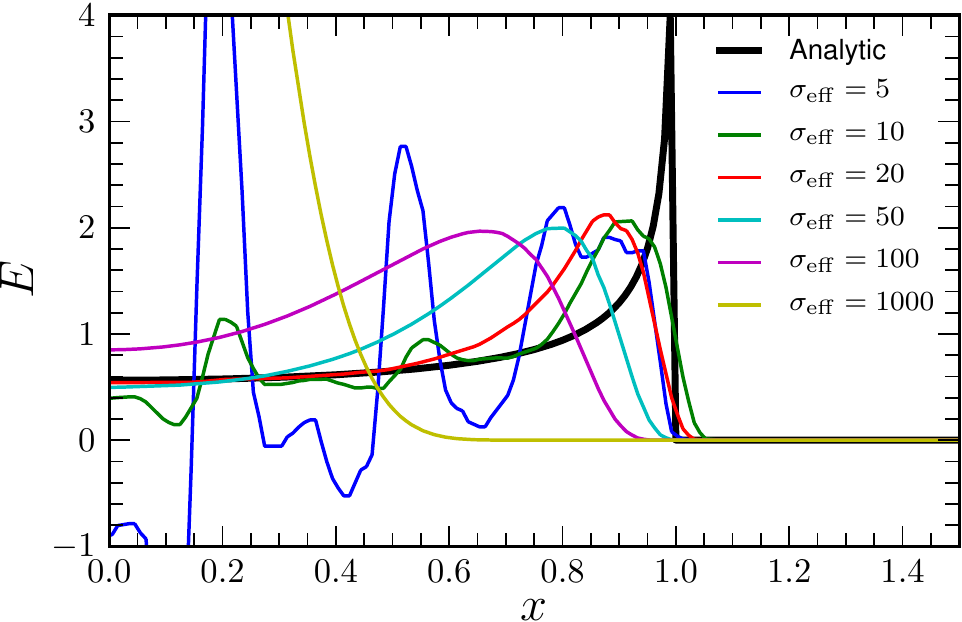}
    \hskip 0.5cm
    \includegraphics[width=7.75cm,height=5.5cm]{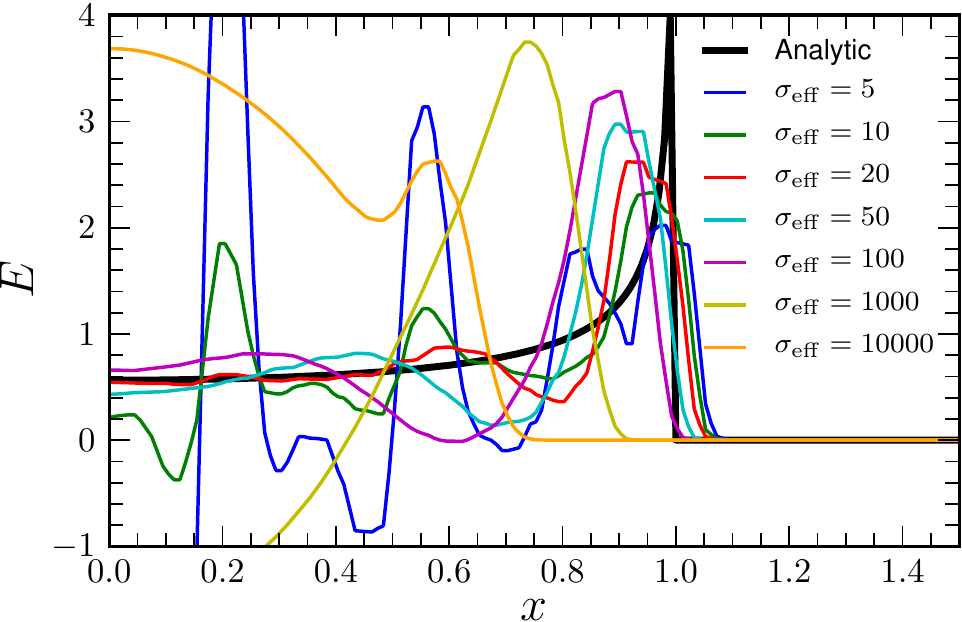}
  \end{center}
  \caption{\textit{Left panel:} radiation energy density as a function of
  the $x$ coordinate at $t=1/c$ obtained from the $F\!P_7$ solution with
  the second-order Lanczos filter of varying strength $\sigma_{\rm eff}$.
  The {thick} black line corresponds to the analytic solution, while the
  rest of the lines represent the numerical solutions corresponding to
  different $\sigma_{\rm eff}$.  \textit{Right panel:} same as in the
  left panel but for the $F\!P_7$ solution with the fourth-order
  spherical-spline filter of varying strength $\sigma_{\rm eff}$.}
  \label{fig:linesource1d_filterstrength}
\end{figure}

The left panel of Fig.~\ref{fig:linesource1d_filterstrength}, which
reports the radiation energy density as a function of the $x$ coordinate
at $t=1/c$, highlights how the quality of the solution varies with the
(Lanczos) filter effective opacity. The black line again corresponds to
the analytic solution, while the rest of the lines represent the $F\!P_7$
solution with the Lanczos filter of varying strength $\sigma_{\rm
  eff}$. In the case of weak filters (\eg, $\sigma_{\rm eff} = 1$ or
$\sigma_{\rm eff}=5$), there are significant oscillations, whose
amplitude is significantly reduced as we increase the filter
strength. For example, for $\sigma_{\rm eff}=20$, there are tiny
oscillations, while for $\sigma_{\rm eff}=50$ there are no noticeable
oscillations. However, the quality of the solution with $\sigma_{\rm
  eff}=50$ is actually worse than the one with $\sigma_{\rm eff} = 20$
(\eg, the radiation wavefront lags significantly behind the real
solution) as a result of the large smearing of the radiation beam
produced by the excessive filtering (this is even more evident for
$\sigma_{\rm eff} = 100$ or $\sigma_{\rm eff}=1000$). Therefore, the
filter strength needs to be chosen large enough to damp oscillations and
small enough to avoid excessive smearing of the solution. We have
repeated these runs with the second-order ErfcLog filter and we again get
similar results.

The left panel of Fig. ~\ref{fig:linesource1d_filterstrength} should be
contrasted with the right one, where we show a study of the effect of
different filter opacities for the same problem, but using the
spherical-spline filter. The first thing to notice is that the
spherical-spline filter is never able to completely remove the ``ring''
structure in the solution, even when the filter strength is so strong
that the result resembles the solution of the diffusion equation for this
problem. Secondly, the dependence of the filter behavior on the filter
strength does not seem to be easily predictable: at first, as we increase
filter strength, negative solution disappear (for $\sigma_{\rm eff}
\lesssim 100$), then they reappear for higher values of $\sigma_{\rm
eff}$ around $1000$. We have repeated these runs with the fourth-order
ErfcLog filter and we obtained similar results.

\begin{figure}
  \begin{center}
    \includegraphics{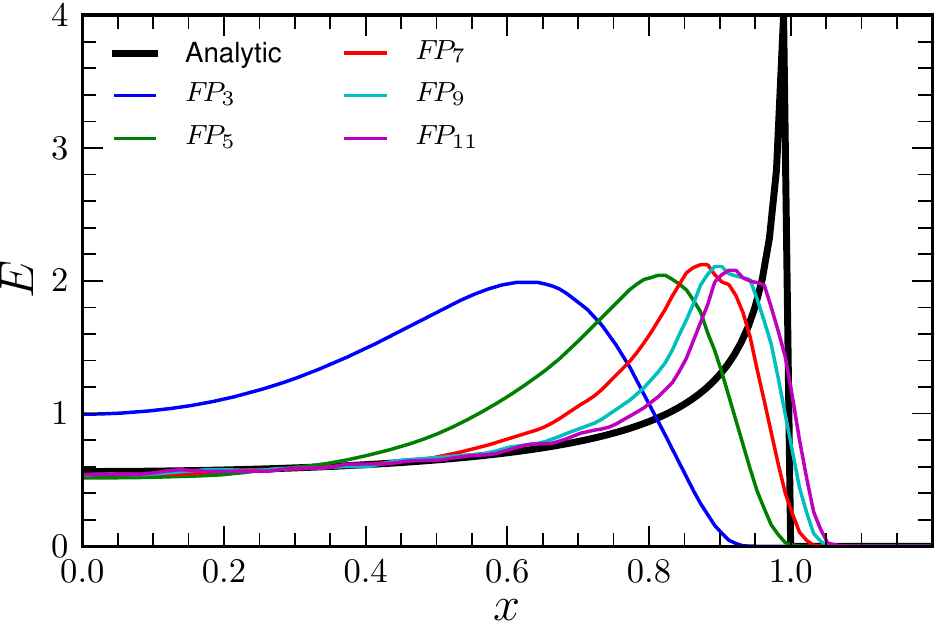}
  \end{center}
  \caption{Radiation energy density as a function of the $x$ coordinate
    at $t=1/c$ obtained from the $F\!P_N$ solution with the second-order
    Lanczos filter with $\sigma_{\rm eff}=20$ for different values of
    order $N$. The {thick} black line represents that analytical
    solution, while the rest of the lines represent numerical solutions
    for different $N$. Clearly, the $F\!P_N$ solution approaches the
    analytical one as we increase the order $N$.}
  \label{fig:linesource1d_order}
\end{figure}

Finally, Fig.~\ref{fig:linesource1d_order} shows the $F\!P_N$ solutions
for $N=3,5,7,9$ and $11$ with the Lanczos filter with $\sigma_{\rm
eff}=20$, and can be used to study the convergence of the $F\!P_N$
approximation to the analytic solution. In this plot, we can distinguish
three different types of errors: (1) An error in the position of the
radiation front, which is particularly evident for small $N$, is mainly
related to the fact that the propagation speed of radiation is smaller
than $c$ for low $N$. (2) An error in the profile of the radiation energy
density behind the front, which is again particularly pronounced for
small $N$, and is due to the fact that high angular resolution is needed
to properly describe the very forward-peaked angular distribution of
radiation. (3) A relatively large spreading of the radiation beam in
space compared to the {$x^{-1/2}$, $x = ct - r$,} singularity in the
analytical solution~(\ref{eq:line_problem_solution}). This is an artifact
of spatial discretization and mainly stems from the fact that the
radiation beam originates from one spatial element and results in the
presence of a ``precursor'' in the radiation front for high-order
$F\!P_N$ (\eg, $F\!P_{11}$) solutions, where the spatial discretization
scheme propagates the radiation front superluminally, despite the fact
that the characteristic speeds of the $F\!P_N$ system are always smaller
than $c$. Superluminal propagation of sharp features in numerical
solutions of hyperbolic PDEs is a well known artifact of the spatial and
temporal discretization of the equations. An in-depth explanation of this
phenomenon for the case of the Maxwell equations can be found in
\cite{Taflove_Hagness2005}. As we can see in the figure, the $F\!P_N$
solution nevertheless approaches the analytical one as we increase the
order $N$ and, in particular, the errors associated with the angular
discretization decrease to the point that the $F\!P_{11}$ solution yields
only a small improvement with respect to the $F\!P_9$ one. In particular,
a large contribution to the error in the $F\!P_{11}$ solution comes from
the presence of the superluminal precursor discussed above. Since this
precursor can only be attributed to the spatial discretization error, we
can conclude that, at this particular resolution and order, the spatial
discretization error is already comparable with the angular
discretization ones. We have repeated these runs with the second-order
ErfcLog filter and at half of the resolution and we again get similar
results. These results show that our filtering strategy is able to
recover the convergence of the $P_N$ scheme for this particular case, and
that the second-order filters, unlike the fourth-order ones, do not
require a delicate fine-tuning of the effective scattering opacity. In
particular, $\sigma_{\rm eff}$ can be chosen on the basis of the physics
and geometry of the problem, in a way that is independent of the order of
the employed $P_N$ scheme.

Overall, the results of this test confirm that the filtering approach is
effective and robust in suppressing unphysical oscillations in the $P_N$
solution, even with moderately low-order $N$. Moreover, we find that the
second-order Lancsoz and ErfcLog filters produce significantly better
numerical results compared to the fourth-order spherical-spline and
ErfcLog filters.

\begin{figure}
  \begin{minipage}{0.49\hsize}
    \center{
      \hspace{-37pt}
      \includegraphics[width=0.64\textwidth]{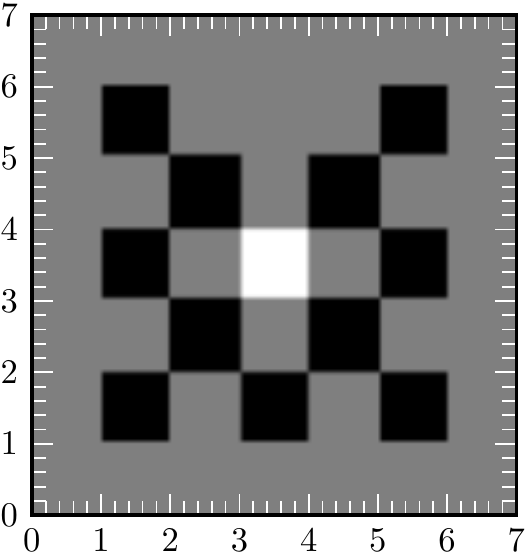}
    }
    \center{\vskip -0.25cm (a) Grid}
    \vskip 0.75cm
    \center{
      \includegraphics[width=0.8\textwidth]{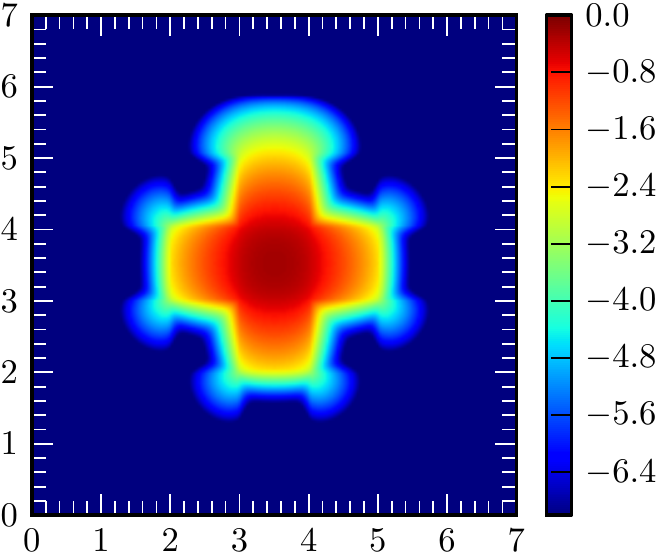}
    }
    \center{\vskip -0.25cm (c) $F\!P_{1}$ solution}
    \vskip 0.75cm
    \center{
      \includegraphics[width=0.8\textwidth]{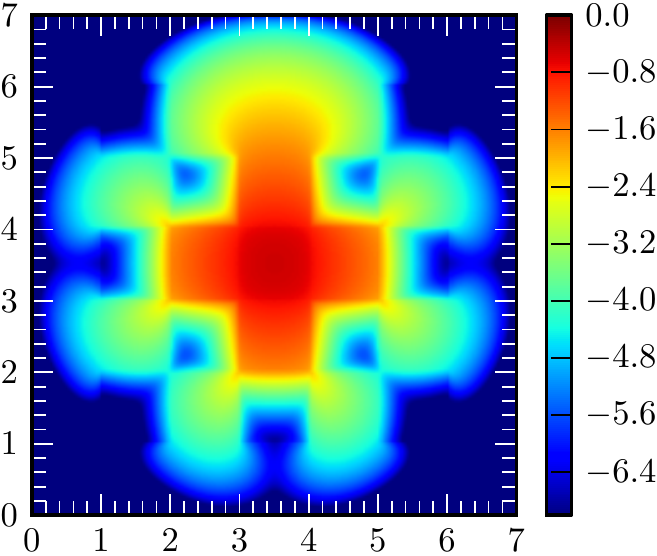}
    }
    \center{\vskip -0.25cm (e) $F\!P_{5}$ solution}
  \end{minipage}
  \begin{minipage}{0.49\hsize}
    \center{
      \includegraphics[width=0.8\textwidth]{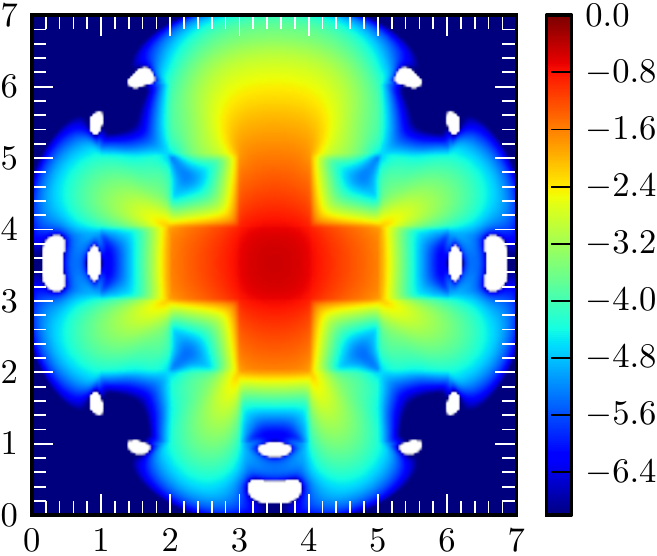}
    }
    \center{\vskip -0.25cm (b) $P_{7}$ solution}
    \vskip 0.75cm
    \center{
      \includegraphics[width=0.8\textwidth]{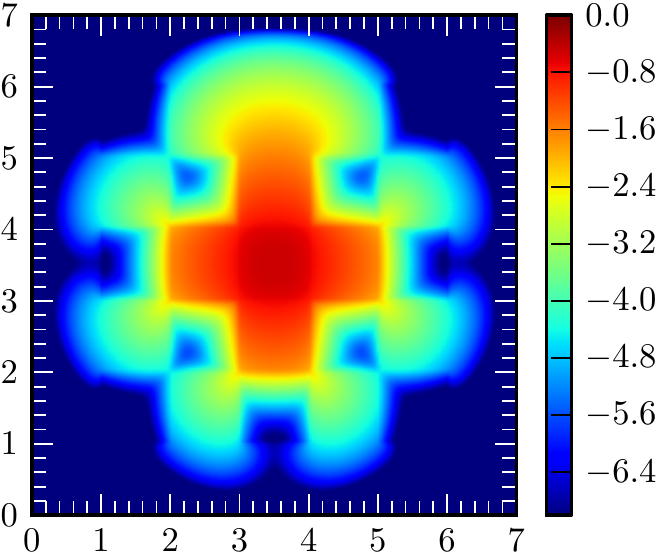}
    }
    \center{\vskip -0.25cm (d) $F\!P_{3}$ solution}
    \vskip 0.75cm
    \center{
      \includegraphics[width=0.8\textwidth]{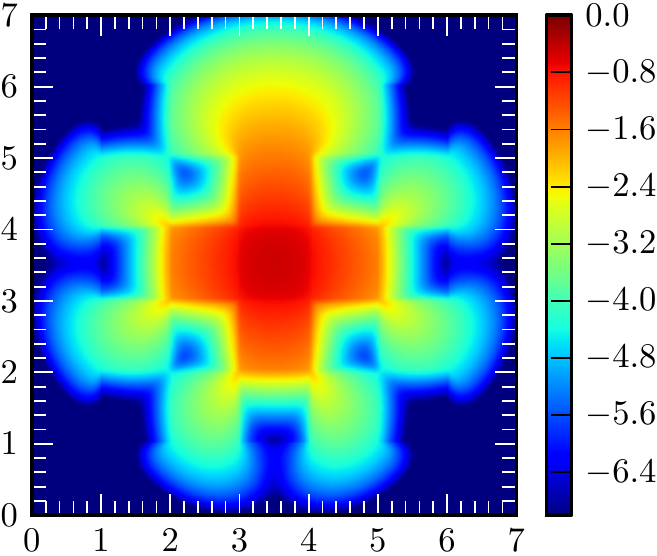}
    }
    \center{\vskip -0.25cm (f) $F\!P_{7}$ solution}
  \end{minipage}
  \begin{center}
    \caption{Upper-left panel: illustration of the setup of the 2D
    lattice problem. The rest of the panels: colormap of the $\log_{10}$
    of the radiation energy density at time $t=3.2/c$ as obtained with
    the $P_7$ scheme (upper-right panel) and $F\!P_N$ schemes with the
    Lanczos filter with opacity 5 for different order $N$ (middle and
    bottom panels). The time $t=3.2/c$ corresponds to the moment when the
    radiation front first reaches the outer boundary of the computational
    domain.}
    \label{fig:2dlattice_test}
  \end{center}
\end{figure}

\subsection{A lattice problem}

Next, we consider another 2D problem consisting of a chessboard of
highly scattering and highly absorbing square regions located around a
central emitting square region. Although this geometry is not expected to
be present in the astrophysical scenarios we are most interested in, it
nevertheless represents a very demanding test of the capabilities of the
different numerical schemes in complicated geometries.

In our calculation we use a setup illustrated in the upper left panel of
Fig.~\ref{fig:2dlattice_test}, which consists of a central emitting
square (shown in white) and $11$ absorbing squares (shown in black) with
a constant absorption opacity $\kappa_a=10$ surrounding the central
emitting square. The space between the squares (shown in gray) and the
central emitting region {have} a small uniform scattering opacity of
$\kappa_s=1$. Each square has a linear size of $1$. The size of the
computational domain is $7$ along both axes. We choose a spatial
resolution of $\Delta x = \Delta z = 0.035$ and the CFL factor was set to
$\approx 0.14$, with outgoing boundary conditions imposed at the outer
boundary.

The remaining panels in Fig.~\ref{fig:2dlattice_test} show the
$\log_{10}$ of the radiation energy density produced by different schemes
at a time $t=3.2/c$, which roughly corresponds to the moment when the
radiation reaches the outer boundary of the computational domain. The
upper right panel corresponds to the $P_7$ solution. Not surprisingly,
this solution has regions of negative energy density (shown in white),
although the negative values reach at most a relatively small magnitude
of $\sim 10^{-5}$. All other solutions are computed with the Lanczos
filter with effective opacity $\sigma_{\rm eff} = 5$. We find that
$\sigma_{\rm eff}\gtrsim 5$ is necessary to avoid the appearance of
negative solutions. The middle-left panel represents the $F\!P_1$
solution, which does not have negative regions, but where the radiation
wavefront has reached only half of the computational domain. This is
again due to the fact that the $N=1$ wave travels at a slower velocity of
$\simeq c/\sqrt{3}$~\cite{Olson00} (see the discussion in
Section~\ref{sec:ang_dis}). The middle-right panel shows the $F\!P_3$
solution, which also does not have regions with negative energy density
and where the front has travelled enough to cover $\sim 95\%$ of the
computational domain, as a result of larger propagation speeds with
higher $N$ (of course, the velocity is always bounded by the speed of
light). The $F\! P_5$ solution shown in the lower-left panel is very
similar to the $F\!  P_3$ case, with the only noticeable difference being
the slightly faster propagation velocity in the $F\!  P_5$ case. Finally,
the lower-right panel shows the $F\! P_7$ solution, which looks almost
indistinguishable from the $F\! P_5$ one.

\begin{figure}
  \begin{minipage}{0.49\hsize}
    \center{
      \includegraphics[width=0.8\textwidth]{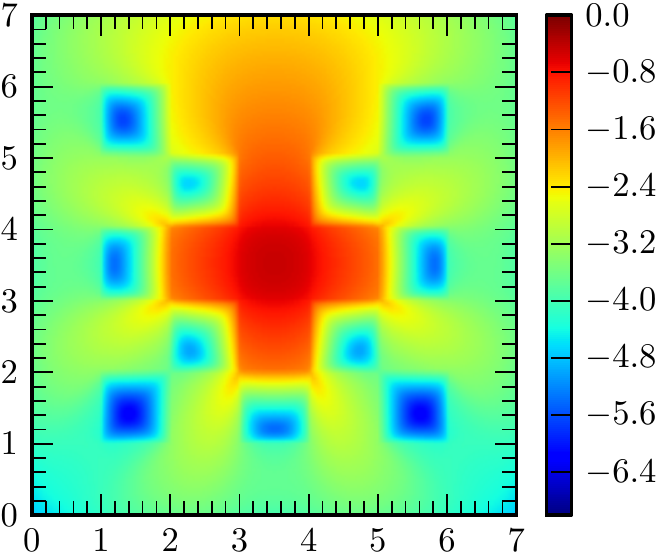}
      \center{\vskip -0.25cm (a) $P_7$ solution}
    }
    \vskip 0.75cm
    \center{
      \includegraphics[width=0.8\textwidth]{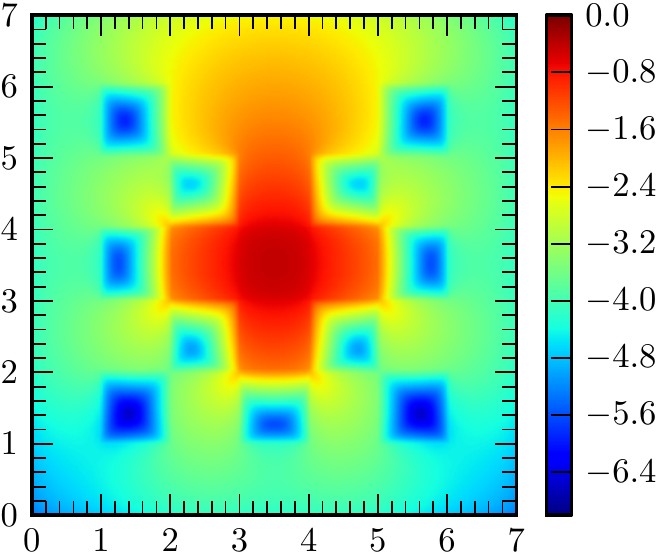}
      \center{\vskip -0.25cm (c) $F\!P_{3}$ solution}
    }
  \end{minipage}
  \begin{minipage}{0.49\hsize}
    \center{
      \includegraphics[width=0.8\textwidth]{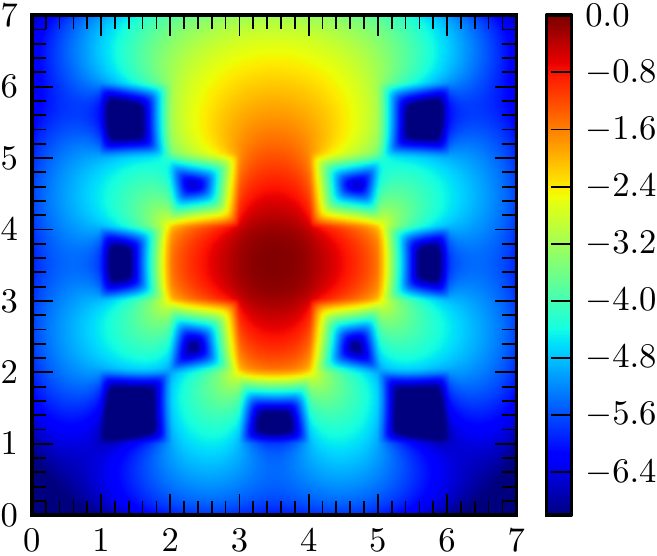}
      \center{\vskip -0.25cm (b) $F\!P_{1}$ solution}
    }
    \vskip 0.75cm
    \center{
      \includegraphics[width=0.8\textwidth]{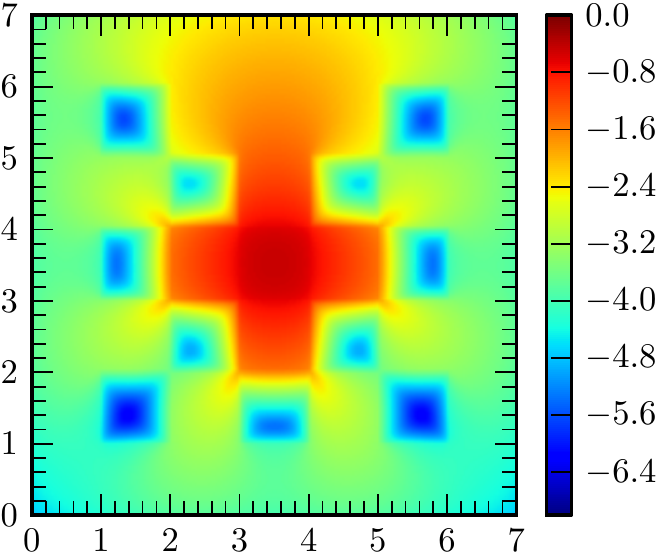}
      \center{\vskip -0.25cm (d) $F\!P_{7}$ solution}
    }
  \end{minipage}
  \caption{Colormaps of the $\log_{10}$ of the radiation energy density
  at time $t=16/c$ obtained with the $P_7$ scheme (upper-left panel) and
  $F\!P_1$ (lower-left panel), $F\!P_7$ (upper-right panel), and $F\!P_3$
  (lower-right panel) schemes with the Lanczos filter with opacity 5. The
  time $t=16/c$ corresponds to the time by which the radiation field
  reaches the stationary state. Since the $P_N$ scheme is less likely to
  exhibit negative solutions in the stationary state, we do not observe
  such solutions in our numerical result (upper-left panel), similarly to
  the filtered $P_N$ solutions.}
    \label{fig:2dlattice_test2}
\end{figure}

We complete the analysis of this test by showing in
Fig.~\ref{fig:2dlattice_test2} equivalent snapshots of the energy density
at a later time of $t=16/c$, when the radiation field has reached a
stationary state. In these conditions, the $P_N$ solution is not expected
to have any negative values \cite{Brunner02}. This is indeed confirmed
by the upper left panel of Fig.~\ref{fig:2dlattice_test2}, which shows
the $P_7$ solution without negative regions. The remaining panels
in Fig.~\ref{fig:2dlattice_test2} report the $F\!P_1$, $F\!P_3$, and
$F\!P_7$ solutions, respectively. Note that the $F\!P_3$, and $F\!P_7$
solutions are very similar to the $P_7$ one, underlining that the filter
we use does not compromise the accuracy of the solution. However, the
$F\!P_1$ solution appears to be significantly different from the $F\!P_3$
and $F\!P_7$ solutions, implying that $N=1$ is not a sufficiently
accurate approximation for this problem.

\begin{figure}
  \begin{center}
    \includegraphics{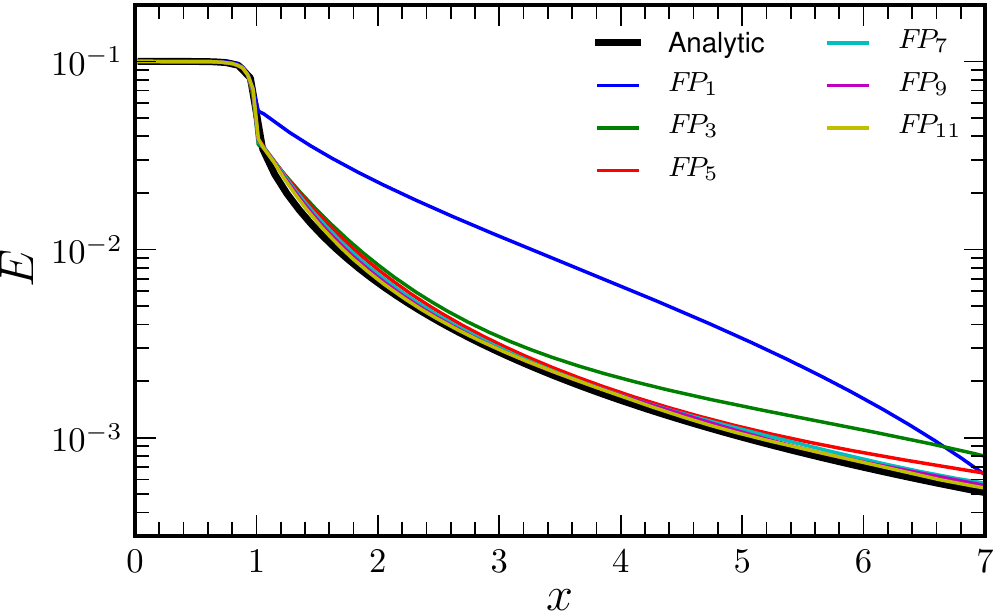}
  \end{center}
  \caption{Radiation energy density as a function of the radial
    coordinate for the homogeneous sphere problem. The {thick} black line
    shows the analytical solution, while the rest of the lines show the
    $F\!P_N$ solution with the Lanczos filter with $\sigma_{\rm eff}=1$
    but with different values of order $N$. Clearly, the numerical result
    approaches the analytical solution as we increase $N$.}
  \label{fig:hom_sphere}
\end{figure}

\subsection{3D Homogeneous sphere}

Finally, we consider the {3D} homogeneous sphere problem, which is
frequently employed to test radiation transport codes~\cite{Smit:97,
Rampp:02, Abdikamalov12}. This problem consists of a static homogeneous
and isothermal sphere of radius $R$ that radiates in vacuum. Inside the
sphere, the radiation interacts with the background matter only via
isotropic absorption and thermal emission. Despite the rather simple
setup, the sharp discontinuity at the surface of the sphere is a model
for astrophysical phenomena with rapidly varying opacity. This represent
a major challenge for finite-difference methods (although, it is less
challenging for Monte Carlo methods; see, \eg,~\cite{Abdikamalov12}).

We assume that the sphere of radius $R$ has a constant absorption opacity
$\kappa_a$ and emissivity $B$ in the interior, while in the ambient
vacuum at $r>R$, we have $\kappa_a = B = 0$. For this problem, the
transport equation can be solved analytically and has
solution~\cite{Smit:97}
\begin{equation}
\label{eq:an_sol}
I(r,\mu) = \frac{B}{\kappa_a}\bigg|_{r=0}\left[1-
\exp\left({-\kappa_a s(r,\mu)}\right)\right] \,,
\end{equation}
where $r \equiv \sqrt{x^2+y^2+z^2}$, $\mu \equiv \cos\theta$ and
\begin{equation}
\label{eq:an_sol_s}
s(r,\mu)=\left\{
\begin{array}{ll}
r\mu+Rg(r,\mu) & \mathrm{if} \ \ r<R, \quad -1\le\mu\le 1 \,, \\
& \\
2Rg(r,\mu) & \mathrm{if} \ \ r\ge R, \quad
\sqrt{1-\left({R}/{r}\right)^2} \le\mu\le 1 \,, \\ & \\
0 & \mathrm{otherwise},
\end{array}
\right.
\end{equation}
and
\begin{equation}
\label{eq:an_sol_g}
g(r,\mu) = \sqrt{1-\left(\frac{r}{R}\right)^2 (1-\mu^2)} \,.
\end{equation}
Note that this solution depends only on three parameters: $\kappa_a$,
$R$, and $B$, where the latter acts as a scale factor for the solution.

\begin{figure}
  \begin{center}
    \includegraphics{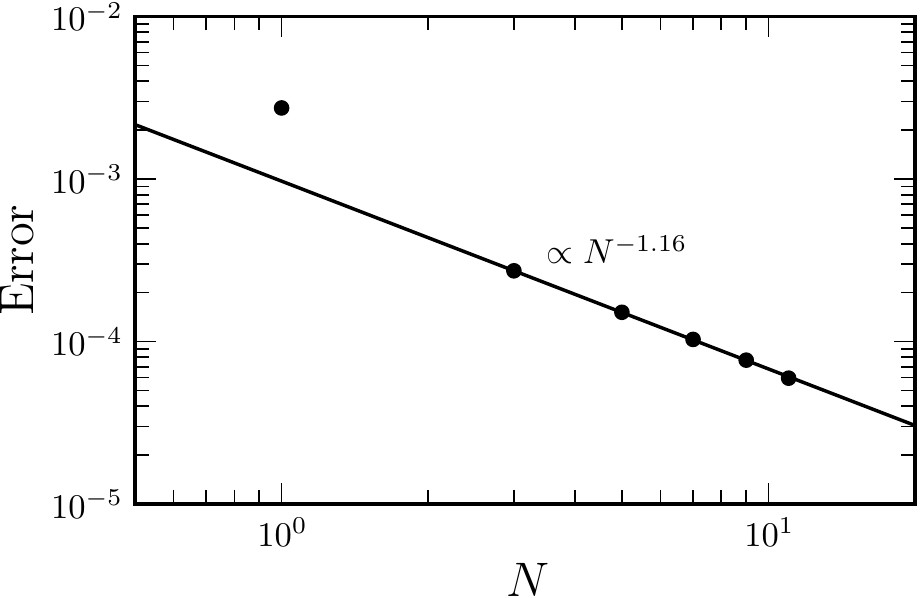}
  \end{center}
  \caption{$L^1-$norm of the deviation of the $F\!P_N$ solution from the
    analytic result as a function of the order $N$ for the homogeneous
    sphere problem. The black dots show the error as computed in a
    {sphere} of radius $R = 4.5$, while the black line shows the fitted
    convergence rate. \label{fig:hsphere.error}
  }
\end{figure}

We perform simulations in full 3D with Cartesian coordinates and use the
following computational setup. We set $R=1$ and cover the interior of the
sphere with $40$ elements in diameter along each coordinate direction,
with the outer boundary being located at $5\,R$. The absorption opacity
is chosen to be $\kappa_a=10$ and the CFL factor is set to be $\sim
0.12$.

It is useful to remark that, although the matter distribution is
spherically symmetric, this is a genuinely 3D test due to the Cartesian
geometry of our spatial grid. Indeed, it leads to the propagation of
radiation from one spatial zone to another not only in the radial
direction, but also in the angular directions, and the degree of
sphericity of the numerical solution can be taken as a measure of the
accuracy.

\begin{figure}
  \begin{center}
    \includegraphics{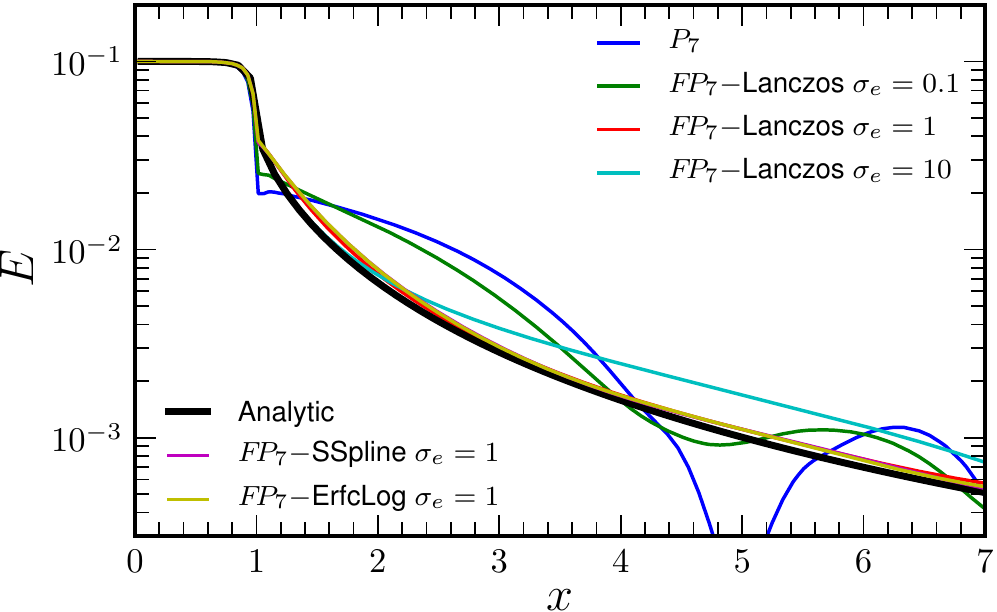}
  \end{center}
  \caption{Radiation energy density as a function of the $r$ coordinate
    for the homogeneous sphere problem. The {thick} black line shows the
    analytical solution, the blue line corresponds to the unfiltered $P_7$
    solution, while the rest of the lines represent the $F\!P_7$
    solutions obtained with different filters and different values of
    $\sigma_{\rm eff}$.}
  \label{fig:hom_sphere_filters}
\end{figure}

Figure~\ref{fig:hom_sphere} shows the radiation energy density along the
diagonal direction for the analytical solution and the $F\!P_N$ solutions
of different orders ranging from $1$ to $11$. The results are shown for
the time when the radiation field has reached a stationary
state.\footnote{Note that stationarity is reached at different times
depending on the different schemes used. For this reason and given the
high computational costs, we did not evolve all the models up to the same
time. Instead we report the solution as obtained as soon as stationarity
is reached. In all cases the computations are performed up to at least $t
= 20/c$.} These runs are performed with the Lanczos filter of
$\sigma_{\mathrm{eff}} = 1$. Interestingly, all of the $F\!P_N$ solutions
produce the correct result in the interior of the sphere. This is not
surprising since the radiation inside the sphere is nearly isotropic and
the low-order $F\!P_N$ solutions are already accurate enough. Outside the
sphere, radiation streams freely outwards with a highly forward-peaked
distribution in angle, which is a challenge for low-order $F\!P_N$
schemes. Indeed, the $F\!P_1$ result deviates significantly from the
analytical solution in that region. However, the solution clearly becomes
more accurate everywhere in the computational domain as we increase the
order of the scheme.  Figure~\ref{fig:hsphere.error} shows the $L^1-$norm
of the deviation of the $F\!P_N$ solution from the analytic result in a
sphere of radius $R=4.5$.\footnote{We compute the error inside $R=4.5$
instead of $R=5$ in order to exclude effects due to boundary conditions.
We also normalize the $L^1-$norm by dividing it by $\frac{4}{3}\pi R^3$.}
As we can see from the plot, the $F\!P_N$ scheme starts to converge
already for $N \approx 3$, an order with only $4^2=16$ angular degrees of
freedom. The convergence order is $\simeq 1.16$, which is consistent with
what expected from the theory of spectral filtering.

Figure~\ref{fig:hom_sphere_filters} shows the radiation energy density
along the diagonal direction for the analytical solution and different
numerical solutions at stationarity state. These are the unfiltered $P_7$
solution, the three $F\!P_7$ solutions computed using respectively the
second-order Lanczos, the fourth-order spherical-spline and ErfcLog
filters with $\sigma_{\rm eff} = 1$.  Also shown are the two $F\!P_7$
solutions computed using the Lanczos filter with $\sigma_{\rm eff} = 0.1$
and $\sigma_{\rm eff} = 10$. Although the $P_7$ solution does not exhibit
any negative solutions, it shows large oscillations in the free streaming
region. The $F\!P_7$ solution with the Lanczos filter with $\sigma_{\rm
eff} = 0.1$ also yields a somewhat oscillatory solution, suggesting that
the filter effective opacity is too low for this problem.  As
$\sigma_{\rm eff}$ is increased, the spurious oscillations disappear and
all the filters that we have tried yield solutions of very similar
quality for $\sigma_{\rm eff} = 1$. Finally, the $F\!P_7$ solution with
the Lanczos filter with $\sigma_{\rm eff} = 10$ is similar in quality to
the $F\!P_3$ solution with the same filter but with $\sigma_{\rm eff} =
1$. This is due to the excessive damping of the high-order multiples of
the solution by the filtering procedure.

\begin{figure}
  \begin{minipage}{0.49\hsize}
    \center{
      \includegraphics[width=0.8\textwidth]{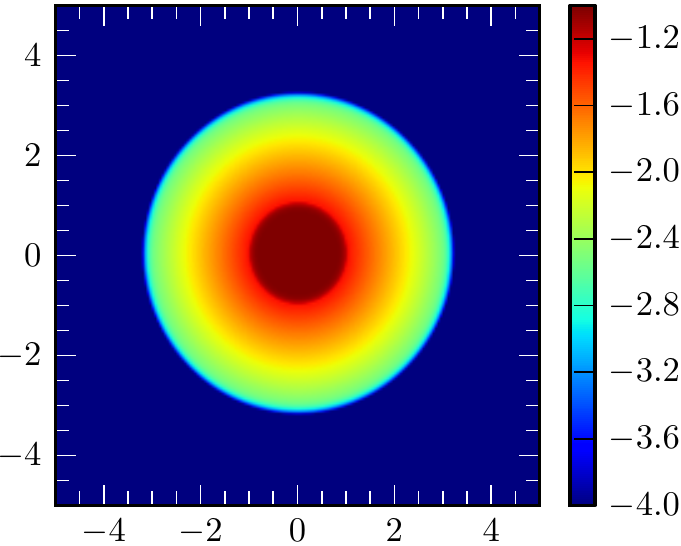}
      \center{(a) $F\!P_1$ solution}
    }
  \end{minipage}
  \begin{minipage}{0.49\hsize}
    \center{
      \includegraphics[width=0.8\textwidth]{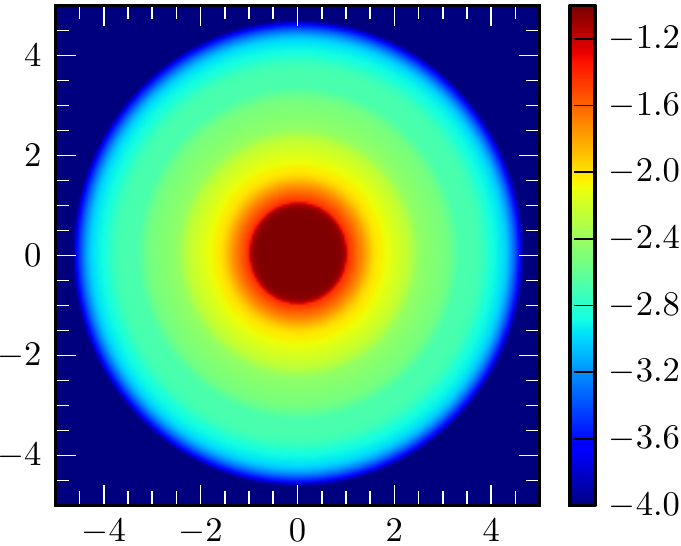}
      \center{(b) $F\!P_{11}$ solution}
      }
  \end{minipage}
  \caption{Colormaps of the radiation energy density at time $t=3.75/c$
    obtained with the $F\!P_1$ and $F\!P_7$ schemes using the Lanczos
    filter with opacity $\sigma_{\rm eff}=1$. The time $t=3.75/c$
    corresponds to the moment when the radiation front almost reaches the
    outer boundary of the computational domain.}
  \label{fig:hom_sphere2}
\end{figure}

Finally, Fig.~\ref{fig:hom_sphere2} shows the colormaps of the
$\log_{10}$ of the radiation energy density from the $F\!P_1$ (left
panel) and $F\!P_{11}$ (right panel) solutions with the Lanczos filter
with $\sigma_{\rm eff}=1$ at $t=3.75/c$, which corresponds to the time
when the radiation front almost reaches the outer boundary. Clearly, and
as observed also in the previous tests, the the $F\!P_1$ radiation front
lags significantly behind the $F\!P_{11}$ one because of its slower
propagation speed (cf. the discussion in
Section~\ref{sec:scheme_filter}). Moreover, we can also see that both
solutions maintain a high level of spherical symmetry despite the
Cartesian geometry of the spatial grid. We obtain similar results in
tests where the sphere was covered by $20$, $10$ and $5$ elements.

\section{Conclusion}
\label{sec:conclusion}

We have presented an extension of the filtered spherical harmonics method
by McClarren and Hauck \cite{McClarren10}, the $F\!P_N$ scheme, to three
dimensions. We have developed the new 3D/multigroup radiation transport
code {\tt Charon}, built within the {\tt Cactus Computational
  Toolkit}~\cite{Goodale2002,cactusweb}. {\tt Charon} uses an
asymptotic-preserving linear discontinuous Galerkin discretization scheme
in space~\cite{McClarren2008} and a semi-implicit time integration
scheme~\cite{McClarren2008a} (cf.  Section~\ref{sec:scheme}).

Our filtering scheme differs from the one presented by \cite{McClarren10}
in one important aspect: we reformulate the filtering procedure so that
it acquires a well-defined continuum limit. In particular, we have shown
that in the limit where the spatial and time steps are reduced to zero,
our filtering scheme can be interpreted as the addition of a
forward-peaked artificial scattering term to the $P_N$ equations. The
filtering procedure is also constructed in such a way as to retain the
convergence of the $F\!P_N$ solution to the solution of the transport
equation as $N\to\infty$.

We have tested our scheme against a few challenging benchmark problems
for radiation transport using four different filtering kernels: the
fourth order spherical-spline filter, which is similar in spirit to the
filter used by \cite{McClarren10}, the fourth-order and second-order
ErfcLog filters~\cite{Boyd1996}, and the classical second-order Lanczos
filter~\cite{Boyd00}. Our findings indicate that the $F\!P_N$ scheme
behaves well also in the three-dimensional case. In addition, we have
shown that the second-order filters are more robust and accurate and
require somewhat less tuning of the filter strength when compared to the
fourth-order spherical-spline and the ErfcLog filters. Since the order of
a filter is one of its most important properties, this result is likely
to apply also to several other second- and fourth-order filters.

In future work, we plan to extend our numerical algorithms to include
velocity dependence, the coupling to hydrodynamics, and, eventually,
general relativity.

\section{Acknowledgments}

We are happy to acknowledge helpful exchanges with Adam Burrows,
Filippo Galeazzi, Cory D. Hauck, Ryan G. McClarren, Christian
Reisswig and Erik Schnetter. This work was supported in part by NSF
under grant nos. AST-0855535, OCI-0905046, by the Sherman Fairchild
Foundation and the Alfred P. Sloan Foundation, by the DFG grant
SFB/Transregio~7, and by ``CompStar'' a Research Networking Programme
of the European Science Foundation. Results presented in this article
were obtained through computations on the AEI computer cluster
``Datura'', on machines of the Louisiana Optical Network Initiative
under grant loni\_numrel07, on the Caltech compute cluster ``Zwicky''
(NSF MRI award No. PHY-0960291), on the NSF XSEDE network under
computer time grant TG-PHY100033, and at the National Energy Research
Scientific Computing Center (NERSC), which is supported by the Office
of Science of the US Department of Energy under contract
DE-AC03-76SF00098. This work was initiated at the MICRA workshop at
the Perimeter Institute in 2011.

\appendix

\section{Real spherical harmonics}\label{appendix:sharmonics}

This Appendix is dedicated to the derivation of the real spherical
harmonics, whose implementation in {\tt Charon} has been particularly
advantageous. We start by recalling that the spherical harmonics,
$Y^m_\ell$ are the eigenfunctions of the Laplace-Beltrami operator,
$\triangle$, on the unit 2-sphere:
\[
  \triangle Y^m_\ell = - \ell (1 + \ell) Y^m_\ell\,.
\]
Spherical harmonics are usually written, in complex form, as
\[
  Y^m_\ell(\varphi,\theta) = e^{im\varphi} P^{m}_\ell(\cos \theta)\,,
\]
where $-\ell \leq m \leq \ell$, $P^m_\ell$ are the associated Legendre
function, see \eg~\cite{Boyd00},
\[
  P^m_\ell(x) = (1-x^2)^{m/2} C^{m+1/2}_{\ell-m}(x), \quad m \geq 0\,,
\]
and $C^\alpha_n$ are the Gegenbauer polynomials of index $\alpha$ and
degree $n$. We, also, use the standard convention that
\begin{equation}\label{eq:alegendre.convention}
  P^{-m}_\ell(x) = (-1)^m\, \frac{(\ell-m)!}{(\ell+m)!}\, P^m_\ell(x)\,.
\end{equation}
The associated Legendre functions of index $m \geq 0$ are orthogonal
in $[-1,1]$ with unit weight \cite{Abram_Stegun1968},
\[
  \int_{-1}^1 P^m_\ell(x)\, P^{m}_{\ell'}(x)\, \dd x =
  \frac{2 (\ell+m)!}{(2\ell + 1) (\ell - m)!}\, \delta_{\ell \ell'} \,,
\]
while the associated Legendre functions of degree $\ell$ and index $m, m' \geq
0$ are orthogonal in $[-1,1]$ with weight $(1-x^2)^{-1}$
\cite{Abram_Stegun1968},
\[
  \int_{-1}^1 P^m_\ell(x)\, P^{m'}_{\ell}(x)\, \frac{\dd x}{1-x^2} =
  \frac{(\ell+m)!}{m(\ell-m)!}\, \delta_{m m'}\,.
\]
The spherical harmonics with index $m, m' \geq 0$ are then orthogonal
with unit weight on the sphere:
\begin{align}
  \int_{\mathcal{S}_1} Y^m_\ell(\varphi, \theta)\, Y^{m'}_{\ell'}(\varphi,
  \theta)\, \dd\Omega
  &= \int_0^{2\pi} e^{i(m-m')\varphi} \dd \varphi \int_0^\pi
  P^m_\ell(\cos \theta)\, P^{m'}_{\ell'}(\cos \theta) \sin \theta \dd
  \theta\\
  &= \int_0^{2\pi} e^{i(m-m')\varphi} \dd \varphi \int_{-1}^1 P^m_{\ell}(x)\,
  P^{m'}_{\ell'}(x)\, \dd x =\\
  &= 2\pi\, \frac{2 (\ell+m)!}{(2\ell + 1) (\ell - m)!}  \, \delta_{m m'}\,
  \delta_{\ell \ell'}\,.
\end{align}
We can then redefine the spherical harmonics as
\[
  Y^m_\ell(\varphi,\theta) =
    \sqrt{\frac{(2 \ell + 1)}{4\pi}\, \frac{(\ell - m)!}{(\ell + m)!}} \,
    e^{i m \varphi}\, P^m_\ell(\cos \theta)
    = N^m_\ell\, e^{i m \varphi}\, P^m_\ell(\cos \theta)\,.
\]
It is easy to see, that for all $m, m'$, thanks to the normalization and the
convention (\ref{eq:alegendre.convention}), we have
\[
  \int_{\mathcal{S}_1} Y^m_\ell(\varphi, \theta)\,
  Y^{m'}_{\ell'}(\varphi,\theta)\, \dd \Omega =
  \delta_{m m'} \delta_{\ell \ell'}\,.
\]
We can construct a real basis from the spherical harmonics by defining
\[
  Y_{\ell m} = \begin{cases}
    \frac{1}{\sqrt{2}} (Y^m_\ell + (-1)^m Y^{-m}_\ell), & \textrm{if } m > 0\,, \\
  Y^0_\ell, & \textrm{if } m = 0\,, \\
    \frac{1}{i\sqrt{2}} (Y^{-m}_\ell - (-1)^m Y^{m}_\ell) & \textrm{if } m < 0\,.
  \end{cases}
\]
Using again the fact that, $N^{-|m|}_\ell P^{-|m|}_\ell = (-1)^{|m|}
N^{|m|}_\ell P^{|m|}_\ell$, we obtain
\begin{align*}
  & Y_{\ell m}(\varphi, \theta) = \sqrt{2}\, N^m_\ell\, \cos(m \varphi)\,
  P^m_\ell(\cos \theta), & & m > 0\,, \\
  & Y_{\ell m}(\varphi, \theta) = \sqrt{2}\, N^{|m|}_\ell\,\,
  \sin(|m| \varphi)\, P^{|m|}_\ell(\cos \theta), & & m < 0,
\end{align*}
which is the wanted expression for the real spherical
harmonics.

We can construct a Gaussian quadrature associated with the spherical
harmonics basis as a direct product of a uniform quadrature in the
$\varphi$ direction:
\begin{align*}
  w = \frac{\pi}{M_\varphi+1/2}, & &
  \varphi_m = \frac{\pi}{M_\varphi+1}\, m, & &
  -M_\varphi \leq m \leq M_\varphi\,.
\end{align*}
which is accurate to order $4 M_\varphi + 2$ \cite{Boyd00} and a
Gauss-Legendre grid in the $\theta$ direction \cite{canuto_2006_smf}:
\begin{align*}
  \mu = \cos\theta,
& &  w_\ell = \frac{2}{(1-\mu_\ell^2)\big[P'_{M_\theta+1}(\mu_\ell)\big]^2},
& &  \mu_\ell = \big\{\mathcal{Z}[P_{M_\theta+1}(\mu)]\big\}_\ell,
& &  0 \leq \ell \leq M_\theta\,,
\end{align*}
where $P_M$ is the Legendre polynomial of degree $M$, \ie $P_M = P^0_M$
and $\mathcal{Z}[p]$ denotes the set of the roots of $p$.
This quadrature formula is then accurate up to order $2M_\theta+1$. This
means that if we want to obtain an exact representation of the scalar
product of spherical harmonics up to order $N_\theta$ we have to choose
$M_\theta = N_\theta$.

\bibliographystyle{plain}
\bibliography{references}

\end{document}